\documentclass[twocolumn,twocolappendix]{aastex63}

\received{Feb 18, 2021}
\revised{Jun 9, 2021}
\accepted{Jun 11, 2021}
\submitjournal{Icarus}

\shorttitle{Convection and cloud formation over the $24\degree$N jet}
\shortauthors{Sankar et al.}
\graphicspath{{./}{figures/}}

\usepackage{amsmath}
\usepackage{gensymb}

\begin{document}

\title{The aftermath of convective events near Jupiter's fastest prograde jet:  implications for clouds, dynamics and vertical wind shear}

\correspondingauthor{Ramanakumar Sankar}
\email{rshankar2012@my.fit.edu}

\author[0000-0002-6794-7587]{Ramanakumar Sankar}

\author[0000-0002-8899-3769]{Chloe Klare}

\author[0000-0001-6111-224X]{Csaba Palotai}
\affiliation{Florida Institute of Technology \\
150 W University Boulevard \\
Melbourne FL, 32901, USA}




\begin{abstract}
The $24\degree$ N jet borders the North Tropical Belt and North Tropical Zone, and is the fastest prograde jet on Jupiter, reaching speeds above $170$ m/s. In this region, observations have shown several periodic convective plumes, likely from latent heat release from water condensation, which affect the cloud and zonal wind structure of the jet. We model this region with the Explicit Planetary hybrid-Isentropic Coordinate model using its active microphysics scheme to study the phenomenology of water and ammonia clouds within the jet region. On perturbing the atmosphere, we find that an upper tropospheric wave develops that directly influences the cloud structure within the jet. This wave travels at $\sim75$ m/s in our model, and leads to periodic chevron-shaped features in the ammonia cloud deck. These features travel with the wave speed, and are subsequently much slower than the zonal wind at the cloud deck. The cloud structure, and the slower drift rate, were both observed following the convective outbreak in this region in 2016 and 2020. We find that an upper level circulation is responsible for these cloud features in the aftermath of the convective outbursts. 
The comparatively slower observed drift rates of these features, relative to the wind speed of the jet,  provides constraints on the vertical wind shear above the cloud tops, and we suggest that wind velocities determined from cloud tracking should correspond to a different altitude compared to the $680$ hPa pressure level. 
We also diagnose the convective potential of the atmosphere due to water condensation, and find that it is strongly coupled to the wave.
\end{abstract}



\section{Introduction and motivation}
Studies of the jovian atmosphere have demonstrated the importance of convective upwelling, with regards to the global energy budget and atmospheric flow \citep{Gierasch2000,Ingersoll2000,Showman2007}. 
These upwellings result in a myriad of changes to the cloud formation in the latitudinal bands where they occur, through modifying the wind, thermal, chromophoric and condensible structure of the atmosphere in their vicinity \citep{SanchezLavega2008, SanchezLavega2017, Fletcher2017, Fletcher2017b, Rogers2019, PerezHoyos2020}. The large scale cloud features that form are transient, and markedly different from those that form during the quiescent state. To explain these features and study their effects, it is necessary to resolve the coupling between the atmospheric dynamics and the physics of clouds formation. In this study, we use a numerical model with cloud microphysics to simulate the jovian atmosphere in the wake of these convective eruptions, in an effort to explain the processes that drive these features and their implications on the local atmospheric structure and dynamics. 

\begin{figure}
    \centering
	\includegraphics[width=\columnwidth]{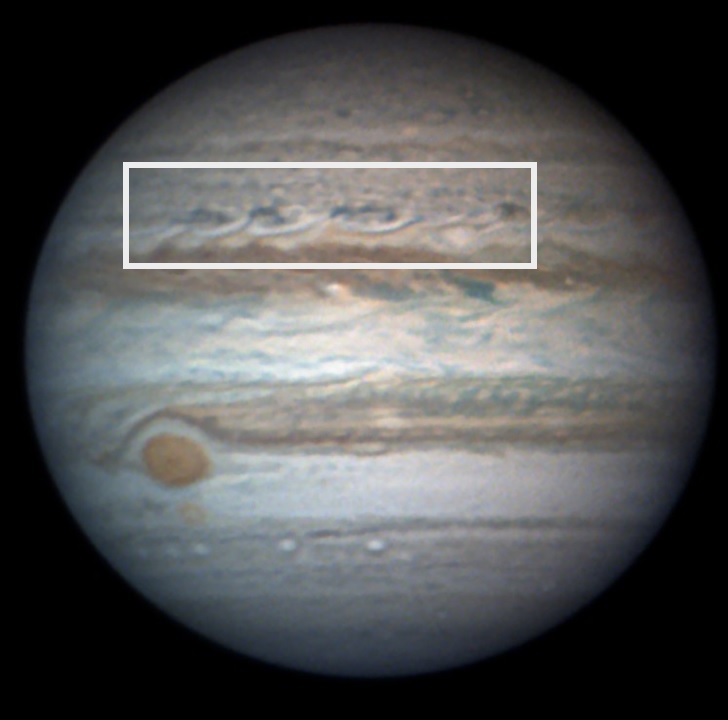}
    \caption{Observations of alternative bright and dark cloud pattern in the jet after an outbreak in 2016 \citep{SanchezLavega2017}. Credit: L~Dauvergne/Fran\c{c}ois~Colas/S2P/IMCCE/OMP}
    \label{fig:wave_obs}
\end{figure}

\begin{figure*}
    \centering
	\includegraphics[width=0.8\textwidth]{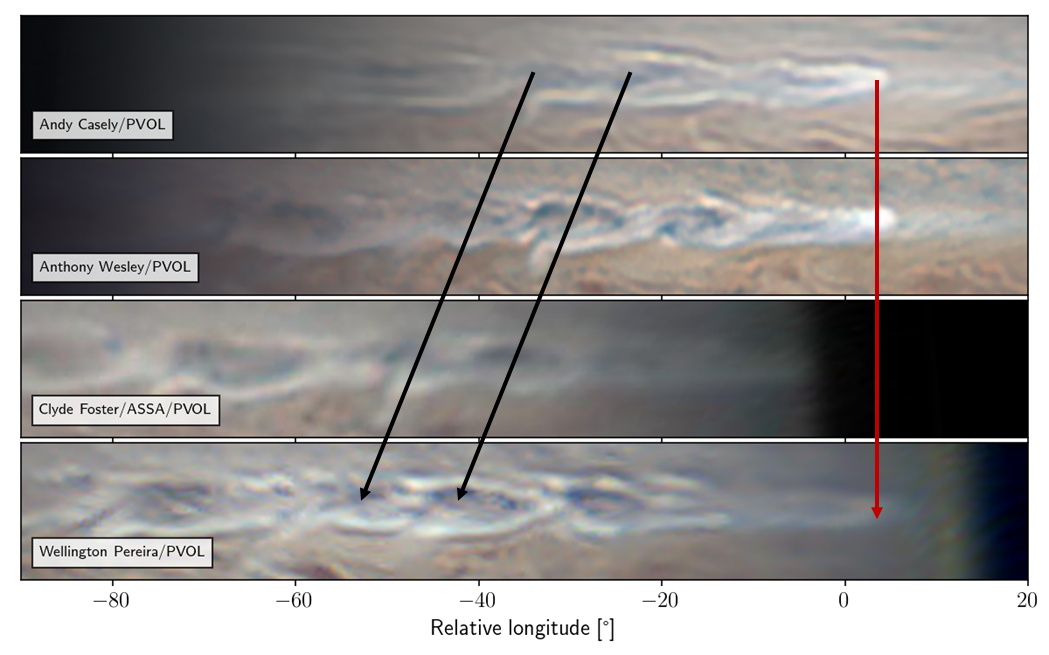}
    \caption{Observations of the plume oubreak in October 2020. The four panels track the head of the plume at different times, as it travels at the velocity at the jet peak of $\sim 150$ m/s. The chevron shape features in the wake fall behind due to their relatively slower speed. }
    \label{fig:plume_obs}
\end{figure*}

Recent observations at a planetographic latitude of $\sim 24\degree$ N (near the fastest eastward jet on Jupiter) show that these upwellings occur every 4-5 years \citep{Fletcher2017b, Rogers2019}. 
Figures~\ref{fig:wave_obs} and~\ref{fig:plume_obs} show the the cloud features associated with such outbreaks. These wave-like features usually have a speed less than that of the jet (Figure~\ref{fig:plume_obs}), and contain an alternating pattern of bright clouds and dark, cloud-free regions \citep{SanchezLavega2017}. This wave is therefore strongly correlated with the cloud formation within the jet, and so is key in understanding the cloud structure and convective outbreaks in this region.

The driver for the upwelling is most likely a result of moist convective water storms, where the aerosols are carried upward from the strong latent heat release from the deep water clouds \citep{Stoker1986, Fletcher2017}. Therefore, studying these storms is vital in understanding the structure of the deep atmosphere of Jupiter and their effects on the upper level clouds. 

In this paper, we present our modeling work with the Explicit Planetary hybrid-Isentropic Coordinate (EPIC) General Circulation Model (GCM) \citep{Dowling2006}. We study the effect of varying the amount of water and ammonia in the atmosphere, and the subsequent effect on the dynamics of the region. These species condense directly from vapour, unlike NH$_4$SH, which we ignore due to it forming from a chemical reaction between ammonia and  hydrogen sulfide. NH$_4$SH also generally does not form thick cumulus clouds, and therefore, is unlikely to produce these convective upwellings \citep{Carlson1988}. Our model domain is restricted to the North Equatorial Belt (NEB), North Tropical Zone (NTZ) and North Tropical Belt (NTB), i.e. $\sim 15\degree-33\degree$~N planetographic latitude on Jupiter.

Several studies have previously modeled jovian atmospheric dynamics and cloud formation to study the 3-dimensional atmospheric structure in this region. Of note due to the similarity to this work is the attempt of \citet{SanchezLavega2008}, who used a dry version of EPIC model to study the outbreak near the NTB, and also a small scale non-hydrostatic model  \citep[based on][]{Hueso2002} to fully simulate the plume outbreak from water condensation. Their work was vital in constraining the vertical wind shear and the deep abundance of water necessary for the formation of these plumes. \citet{SanchezLavega2017} again used the EPIC model to study the dry atmospheric dynamics by triggering outbreaks in the jet as vortices at key locations. This was in an effort to reproduce observations of the pattern of clouds observed in the wake of the plumes. 

Our work aims to augment these prior attempts by fully studying the direct coupling of atmospheric dynamics with the explicit treatment of cloud formation. Latent heat release, especially from water, is the primary driver for these plumes, and thus, has a direct impact on the convective ability in the region.  The goal of this study is to understand the effect of deep abundance of water and ammonia on the convective capability and dynamic structure of the atmosphere. We focus on the 24$\degree$ N jet region to study the effects of the convective outbursts and to reproduce the observed cloud structure following the events. 
In this paper, we present our results showing the persistent wave that forms in this region, following these outbursts, and its effect on the formation of both water and ammonia clouds. We also study the relation between the wave and convective capability, and how this changes with volatile abundance.


In Section~\ref{sec:method}, we describe our model, the initial conditions and the parameter space. We also describe our numerical parameters, and a method for diagnosing convective potential in our hydrostatic model. In Section~\ref{sec:results}, we discuss the results from our model runs, and detail the differences due to our parameter space. In Section~\ref{sec:discussion}, we present a discussion of our results and final thoughts. 

\section{Method}
\label{sec:method}
\subsection{EPIC model and treatment of cloud formation}
To study the jet region, we use the EPIC General Circulation Model (GCM). The model uses a hybrid coordinate system where the top of the model is defined in potential temperature ($\theta$) and the bottom of the model (where $\theta$ is roughly constant) is defined by a scaled pressure ($\sigma$). A generalized coordinate $\zeta$ is defined as a function of the $\sigma$ and $\theta$ to allow for a smooth transition between the two regimes \citep{Dowling2006}.

The cloud formation is modeled as described in \citet{Palotai2008}, which is a single-moment bulk cloud microphysics scheme. The scheme deals with five distinct phases for each condensing species: vapour, cloud ice, cloud liquid, snow and rain. In our simulations, we use both water and ammonia as condensibles, with both treated as pure substances (no reactionary chemistry involved). 
Terminal velocity is parameterized as a power law in diameter and pressure, 

\begin{equation}
    V_{\text{T}} = x\left(\dfrac{D}{1 \text{ m}}\right)^y \left(\dfrac{p_0}{p}\right)^{\gamma},
\end{equation}
where $x$, $y$ and $\gamma$ are the fit parameters, $D$ and $p$ are the particle diameter and atmospheric pressure, respectively, and $p_0 = 1$ bar is the reference pressure. This power law is fit to an explicit formulation of terminal velocity at different Reynolds and Best numbers as detailed in \citet{PruppacherKlettbookCh10}. 


\begin{figure}
	\centering
	\includegraphics[width=\columnwidth]{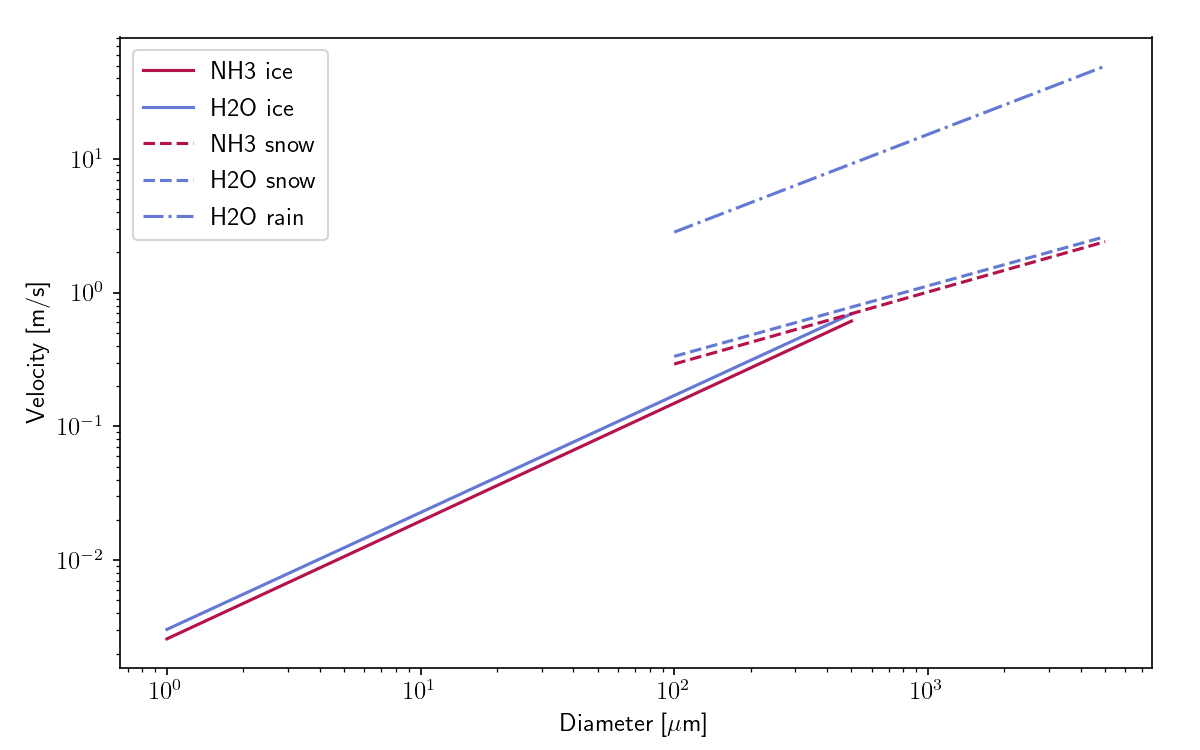}
    \caption{Terminal velocity profiles for ice (solid), snow (dashed) and rain (dot-dashed) in our model. These plots are generated from the fits in Table~\ref{tab:tvelfits}. }
    \label{fig:termvel}
\end{figure}

We have updated our terminal velocity fits to account for ice cloud particles. We assume that cloud particles have a constant mean radius within a grid cell that is a function of the cloud density, using the empirical relation as given in \citet{HDC2004}. Consequently, we determine the terminal velocity assuming that they are porous spheres, with a density half that of bulk ice. Given the small particle sizes of ice compared to snow, we have also updated our fits for all phases to account for the Knudsen number at lower particle sizes \citep{Cunningham1910}, and these parameters for Jupiter are given in Table~\ref{tab:tvelfits} and are plotted in Figure~\ref{fig:termvel}.


\begin{table}
    \caption{Terminal velocity fits of cloud ice, snow and rain for water and and cloud ice and snow for ammonia on Jupiter. Ammonia rain does not form in our model due to the cold temperatures. \label{tab:tvelfits}}
	\begin{tabular}{cccc}
    	\hline
    	& $x$ [m/s] & $y$ & $\gamma$ \\
    	\hline
    	NH$_3$ ice  & $495.22$  & $0.8807$   & $0.1248$ \\
    	NH$_3$ snow & $41.74$   & $0.5386$   & $0.3008$ \\
    	H$_2$O ice  & $535.41$  & $0.8746$   & $0.1264$ \\
    	H$_2$O snow & $42.83$   & $0.5271$   & $0.3043$ \\
    	H$_2$O rain & $2376.36$ & $0.7308$   & $0.3568$ \\
    	\hline
	\end{tabular}
\end{table}

\subsection{Model parameters}

\begin{figure}
	\centering
	\includegraphics[width=\columnwidth]{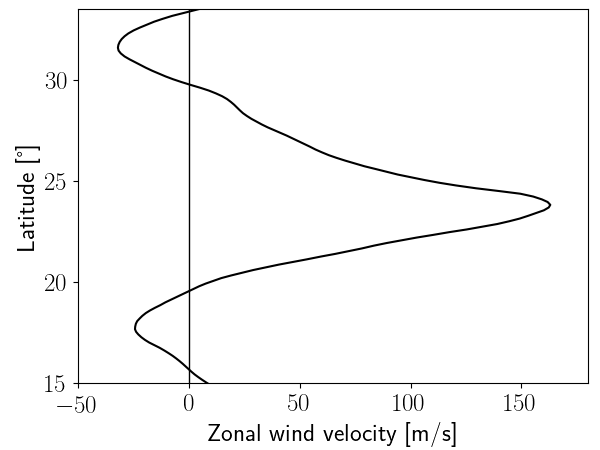}
    \caption{Zonal wind profile used to initialize the model. This wind profile  applied at a pressure of $680$ hPa in our model assuming no wind shear below and wind decay over 2.4 scale-height above.}
    \label{fig:zonal_wind}
\end{figure}

The model extends from $15\degree$ and $33.5\degree$N in latitude with 80 gridpoints, and from $0\degree$ to $120\degree$ in longitude with 256 gridpoints. This gives a resolution of about $0.47\degree$ and $0.23\degree$ in the zonal and meridional directions respectively. The meridional resolution is higher to compensate for geometric effects so as to keep the physical grid size roughly equal in both directions. The boundaries are fixed in the meridional direction and periodic in the zonal direction. This region covers the eastward jet at $23.7\degree$ N and also the two westward jets to the north and south of it.  The zonal wind profile used is from {\it Voyager} data from \citet{Limaye1986} and is shown in Figure~\ref{fig:zonal_wind} for the model region. We make the assumption of no vertical wind shear with depth below $680$ hPa, similar to the results of \citet{SanchezLavega2008,SanchezLavega2017}. Above this, the wind decays to zero in $2.4$ scale heights as determined by \citet{Gierasch1986}, similar to \citet{Palotai2014}.

\begin{figure} 
    \centering
	\includegraphics[width=\columnwidth]{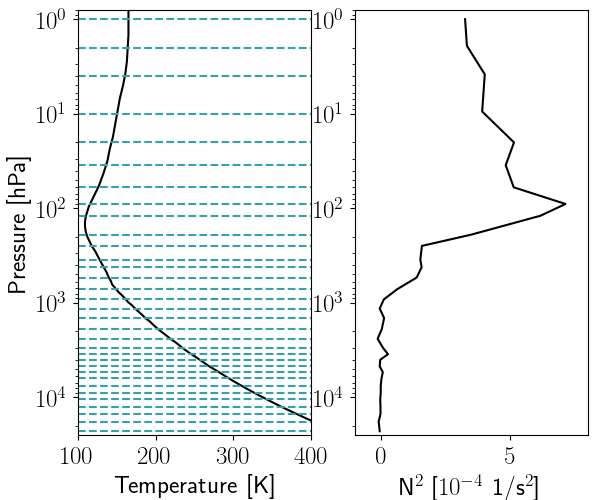}
    \caption{Left: Temperature-pressure of the atmosphere. The dashed blue lines denote model layers. Right: Brunt-V{\"a}is{\"a}l{\"a} frequency of the model atmosphere. Both profiles are for the lower latitude boundary.  }
    \label{fig:tp_bv}
\end{figure}

Vertically, the model ranges from 0.1 hPa to 22800 hPa with 35 total layers. The layers are unevenly spaced to produce higher resolution near features of interest (such as the water and ammonia cloud levels). We place the transition between the $\sigma$ and $\theta$ coordinates at 20 hPa so that the transition does not interfere with the tropopause (at ${\sim}100$ hPa) or the clouds. We use the temperature profile obtained by \citet{Moses2005}.  The temperature profile, vertical layout and the Brunt-V{\"a}is{\"a}l{\"a} frequency is shown in Figure~\ref{fig:tp_bv}. The temperature profile is prescribed on the south boundary, and the thermodynamic equation is used to fill in the values of temperature at other locations, which results in a 2D temperature structure similar to that observed by \citet{Fletcher2020}. A constant timestep of $\Delta t = 36$~s is used throughout each simulation. We use the top four layers as a sponge ($p < 10$ hPa) to dissipate upward travelling waves and apply Rayleigh drag to the bottom four layers ($p > 12.5$ bar) to maintain the deep jet profile. These are placed away from regions of interest (between ${\sim}100-10000$ hPa) so as to ensure that they do not interfere with the validity of the results.

Our version of EPIC features 4th, 6th and 8th order hyperviscosity to dampen high-frequency oscillation. For all the cases, we maintain divergence damping at $\nu_\text{div} = (0.5)[(1/30) \Delta y^2/\Delta t]$ m$^2$/s and 8th order hyperviscosity at $\nu_8 = (0.5)[(1/2400) \Delta y^8/\Delta t]$ m$^8$/s.

\subsection{Deep abundance}
Obtaining tight observational constraints on the deep abundance of volatiles on Jupiter has proved to be challenging, since it is difficult to spectrally identify clouds and determine their densities \citep{Taylor2004,Wong2004PSS,Bjoraker2020}. The {\it Galileo} probe famously descended through a dry region of Jupiter measuring sub-solar abundance of water and roughly $3\times$ solar abundance of ammonia \citep{GalileoProbepaper, Wong2004}. More recently, the {\it Juno} Microwave Radiometer (MWR) has been used to determine a global map of ammonia abundance \citep{Li2017}. The inhomogeniety in ammonia vapour concentration is clear from their results, and  show a deep, well-mixed ammonia layer below 15-20 bars. The ammonia concentration in this deep layer is about $360$ ppm, or roughly $2\times$ the solar value. 

Water has remained even more elusive, with only inferences made of the presence of the deep water cloud \citep[e.g.,][]{Gierasch2000,Bjoraker2015}. {\it Galileo} probe data put a limit on the minimum abundance of ${\sim}0.5\times$ solar \citep{Wong2004}, but there was difficulty putting an upper limit without direct evidence of water vapour spectral lines. Localized results have been obtained by identifying an opaque, optically thick cloud layer below $P > 4$ bars, e.g. under the Great Red Spot \citep{Bjoraker2018}, pointing to a roughly solar water abundance. Near the equatorial regions, \citet{Li2020} use the {\it Juno} MWR instrument to constrain the water abundance to between $1-5\times$ solar. 

Our goal is to investigate the effect of bulk abundance of volatiles on atmospheric stability and convection. In previous studies, the abundance of ammonia has had little effect on atmospheric stability \citep{Palotai2008}. Rather, it is water that contributes to convection given the high latent heat release and high cloud density, both of which the upper tropospheric ammonia clouds lack \citep{Stoker1986}. Consequently, we test values of $0.5\times$, $1\times$ and $2\times$ solar abundance of water in our models. For these three cases, we test $2\times$ and $4\times$ solar ammonia abundance.

\subsection{Balance of the 2D state}
Even though hydrostatic and thermodynamic balance is used to initialize the model grid, numerical artifacts result in spurious unstable growth in the first few timesteps. As such, to ensure that any cloud formation is not a result  of artifacts from model initialization, we run the model without any active microphysics, i.e. in ``passive" mode, where vapour is simply advected and no phase conversions are allowed. We run the model for 200 days to allow these numerical artifacts to stabilize. Furthermore, due to the assumption of zonal symmetry during this phase, we run the model in 2-dimensions (latitude vs. pressure) to reduce the computational cost. At the end of this adjustment, the longitudinal axis is extended by copying the 2-dimensional data along its length. 

\subsection{Breaking the symmetry}
At the end of the spinup phase, the zonal stretching results in a perfect zonal symmetry for the state variables. 
Prior studies have solved this issue by injecting localized heat pulses \citep{GarciaMelendo2005} or through randomized noise in the wind field \citep{Sayanagi2010,MoralesJuberias2011}, which invariably add net energy or vorticity to the model, resulting in an additional free parameter to control in our simulations. Instead, to break this symmetry, we perturb the model by adding random noise to the vorticity field using 100 small vortices distributed throughout the model.
Our noise profiles are created in a way to make sure that the net potential vorticity added to the model as a result of these perturbations is zero. This ensures that we are not affecting the wind structure of the jet by spinning it up/down from the added noise. 
The vortices are introduced as a perturbation to the Montgomery streamfunction, $M = c_p T + gz$, which leads to a change in heat/mass within a grid cell. The perturbation at longitude $\phi$, latitude $\lambda$ and pressure $p$ is defined as a 4-th order polynomial Gaussian ellipsoids,


\begin{equation}
    \Delta M(\phi,\lambda,p) = {\rm amp}_s \times \dfrac{1 + (1-r^4)\exp(2-r^4)}{1 + e^2},
\end{equation}
where ${\rm amp}_s = 1.166 R_e f(\lambda) v_s$ is the amplitude of the perturbation to give a maximum vortex speed of $v_s$ \citep{Stratman2001}, $R_e$ is equatorial radius, $f$ is the Coriolis parameter at that latitude,

\begin{equation}
    r^2 = \dfrac{(\phi-\phi_0)^2}{a^2} + \dfrac{(\lambda-\lambda_0)^2}{b^2} + \dfrac{[\log(p/p_0)]^2}{c^2}
\end{equation}
given a vortex center at $(\phi_0, \lambda_0, p_0)$, and $a, b$ and $c$ are the vortex extents in each direction. The corresponding velocity perturbation is calculated from geostrophic balance from combined effect of the vortices.
In our case, the vortices are defined to be up to $c=1$ scale height vertically, $a=b=1\degree$ horizontally, with velocities $v_s < 10$ m/s, and are placed between $p_0=500$ and $5000$ hPa.

We test three different sets of perturbations that have randomized locations within the model, given the above conditions, for each value of deep abundance. 
These different perturbations are to ensure that our results are consistent despite the profile of the perturbation added. Our final conclusions will be based on an average of these different profiles and will not depend on the exact configuration of this initial condition.
Once the vortices are added, we run the model for 25 days to allow the vortices to adjust. Throughout this phase, we keep the cloud microphysics turned off due to the chaotic nature of the vortex adjustment, as this would lead to unnatural cloud formation. At the end of the 25 days, we trim supersaturated vapor from the model to $100\%$ humidity and turn on the cloud microphysics scheme. This is so that we do not have a sudden burst of cloud formation within the first timestep -- clouds in the model should form in locations driven by the atmospheric flow, rather than through the initialization. The list of test cases is shown in Table~\ref{tab:cases}.

%


\begin{table}
	\caption{List of test cases. Each perturbation number corresponds to a different 
    random vortex profiles}
	\label{tab:cases}
	\begin{tabular}{cccc}
	    \hline
    	Case & H$_2$O & NH$_3$ & Perturbation \\
    	                   &   [solar]        &    [solar]       & \\
    	\hline
    	1 & 0.5 & 2 & 1 \\
    	2 & 1   & 2 & 1 \\
    	3 & 2   & 2 & 1 \\
    	4 & 0.5 & 2 & 2 \\
    	5 & 1   & 2 & 2 \\
    	6 & 2   & 2 & 2 \\
    	7 & 0.5 & 2 & 3 \\
    	8 & 1   & 2 & 3 \\
    	9 & 2   & 2 & 3 \\
    	10 & 0.5 & 4 & 1 \\
    	11 & 1   & 4 & 1 \\
    	12 & 2   & 4 & 1 \\
    	13 & 0.5 & 4 & 2 \\
    	14 & 1   & 4 & 2 \\
    	15 & 2   & 4 & 2 \\
    	17 & 0.5 & 4 & 3 \\
    	18 & 1   & 4 & 3 \\
    	19 & 2   & 4 & 3 \\
    	\hline
    \end{tabular}
\end{table}


\subsection{Convection and CAPE}
Since we are unable to directly resolve convective plumes in EPIC due to its hydrostatic nature, we diagnose the effects of the convection by calculating the convective available potential energy (CAPE) and convective inhibition (CIN). The procedure for these is well formulated for Earth-based meteorology and is used regularly to diagnose storm formation from atmospheric soundings. An example of this calculation can be found in \citet[pp 289-298]{Holtonbook}. For completeness, we describe our formalism using model output in Appendix~\ref{appendix:CAPE}. Using this method, we also calculate the cloud base (or lifting condensation level, LCL) for the parcel. The bottom pressure where the buoyancy of the parcel becomes positive is the level of free convection (LFC) and the top of the convective region, where the buoyancy becomes negative again, is the equilibrium level (EL).  

The CAPE/CIN calculation involves choosing the initial location of the moist parcel, usually below the base of the cloud. This initial parcel conditions on Jupiter is somewhat arbitrary, as even on Earth, the choice of the initial parcel pressure level is controversial \citep{Thompson2007}, and leads to multiple reported values of CAPE. On Jupiter, there is no reference pressure to work with to determine the CAPE, and thus, we determine CAPE starting at pressures of $4$, $5$ and $6$ bars. From our model outputs, we determine a suitable starting pressure as detailed in Section~\ref{sec:results} below.

\section{Results}
\label{sec:results}
Our simulations showed several interesting phenomenology related to cloud formation in this region. Both water and ammonia clouds evolved quickly in the model and reached a quasi-steady state a few days into the simulation. In the following sections, we describe the structure of an upper-tropospheric potential vorticity wave that forms after perturbing the atmosphere in all the cases, and results in distinct large-scale cloud features. 
Hereafter, day zero corresponds to the point where active microphysics is switched on (i.e. 25 days after adding the perturbations; 225 days from the start), and potential vorticity (PV) refers to Ertel's isentropic potential vorticity.  EPIC determines the Ertel PV on on a model layer, as,

\begin{equation}
    q = -g \eta \dfrac{d\zeta}{dp},
\end{equation}
where $\eta$ is the vertical component of the absolute vorticity, and $\zeta$ is the hybrid coordinate. In the top of the model, $\zeta \equiv \theta$, which gives Ertel's PV. Near the bottom, $\zeta = \sigma$, which follows isobaric surface, which makes $q$ proportional to absolute vorticity rather than Ertel's PV on an isentropic surface. We correct for this by recalculating PV on isobaric surfaces, from model output, using the definition,

\begin{equation}
    q = -g \eta \dfrac{d\theta}{dp}.
\end{equation}
Note that Ertel's PV is conserved on an isentrope in the model, internally. The correction here is only for when PV is used as an output. 
Furthermore, all coordinate axis are planetographic and follow Jupiter's System III rotation rate. However, we do note that our longitude system is positive east, for simplicity.

\subsection{Upper level baroclinic wave}
\begin{figure*}
    \centering
    \includegraphics[width=0.8\textwidth]{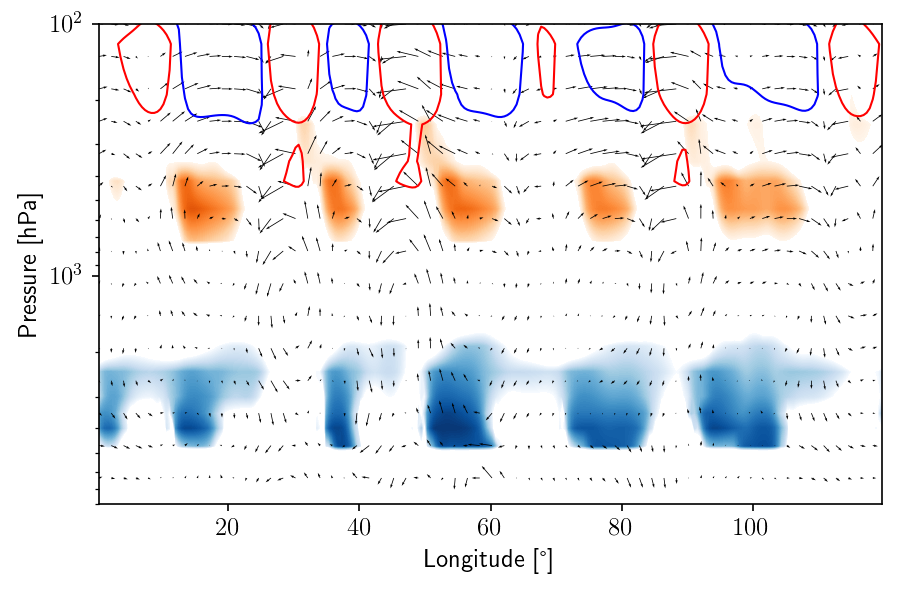}
    \caption{Meridional slice at $23.78\degree$ latitude at day 4 for Case 2. Water clouds are plotted in blue, ammonia in orange. The upper level eddy PV is plotted with blue for positive anomaly (cyclonic) and red for negative (anticyclonic). Wind vectors plot the vertical velocity and anomalous (difference from zonal mean) zonal wind. Vertical velocity is exaggerated by a scale of 40. There is upwelling to the immediate west of the cyclonic packet, resulting in thick cloud formation and downwelling to the east, resulting in cloud clearing. }
    \label{fig:wave_slice}
\end{figure*}
Cloud formation within the jet is strongly influenced by the presence of an upper tropospheric PV wave which forms when the perturbations adjust. 
The waves are the result of a baroclinic instability that occurs within the decaying wind shear above the ammonia cloud top level. The atmosphere is unstable to small perturbations in this region due to the baroclinicity above the cloud tops, as a result of the wind shear resulting in a strong meridional temperature. The sharpness of the jet peak also causes the potential vorticity gradient to change signs within the jet, which is a necessary condition for dynamical instability \citep{CharneyStern}.
A zonal slice at a latitude of $23.78\degree$ is shown for Case 2 in Figure~\ref{fig:wave_slice}. The red contours indicate cyclonic anomaly while the blue indicates anticylonic. The wave packets travel at between 70-80 m/s which is roughly the mean zonal wind speed at 180 hPa -- the vertical center of the wave packets. 

\begin{figure}
    \centering
    \includegraphics[width=\columnwidth]{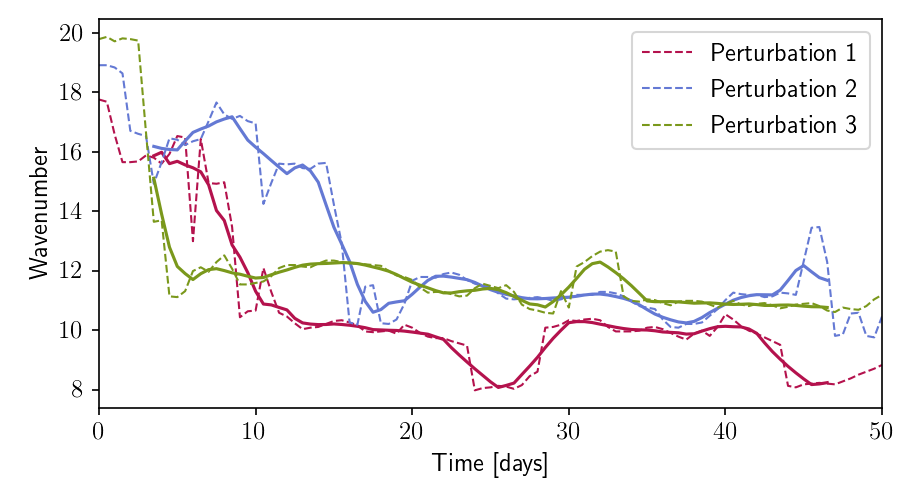}
    \caption{Wavenumber averaged over the different abundances for a given perturbation profile. For both plots, the solid line is the 5-point boxcar average of the raw average (dashed line). All three sets of perturbation demonstrate similar decrease in wavenumber. }
    \label{fig:wave_speed}
\end{figure}

Figure~\ref{fig:wave_speed} shows the wavenumber as a function of time for the different perturbation profiles (averaged over the different initial abundance cases). We determine the wavenumber by calculating a periodogram of the PV at 190 hPa along the $23.7\degree$ latitude transect, and finding the wavelength with the highest power. The wavenumber is then determined as the reciprocal of this peak wavelength. Initially the wavenumber is high (${\sim} 18$), but over time several wave packets dissipate to give a wavenumber of $9-12$, which are stable over the latter half of the simulations. This is true for all simulations, regardless of the perturbation profile.

\subsection{Vertical velocity}
Vertical velocities are calculated diagnostically, from the continuity equation, similar to \citet{Palotai2014},

\begin{equation}
    w = \dfrac{Dz}{Dt} = \dfrac{\partial z}{\partial t} + u \dfrac{\partial z}{\partial x} + v \dfrac{\partial z}{\partial y} + \dot{\zeta} \dfrac{\partial z}{\partial \zeta},
\end{equation}
where $\dot{\zeta}$ is the vertical velocity in the $\zeta$ (i.e, $\sigma$-$\theta$) coordinate and $z$ is the altitude of the layer relative to the base of the model.



The motion of the PV anomaly stretch and compress the deeper atmosphere resulting in up- and downwelling. The region below the cyclonic anomaly is cooler (cold core) and warmer below the anticyclonic anomaly. Thus, the transition from the warmer (anticylonic) to the cooler (cyclonic) region requires upwelling (which cools the upper atmosphere) and the converse requires heating from downwelling. The updrafts lead to an enrichment of volatiles from the deeper regions, forming clouds, while the downdrafts bring dry air from aloft and lead to cloud clearing.


Near the ammonia cloud layer (${\sim}400-600$ hPa), updrafts are between 5-10 cm/s and downdrafts are between 2-5 cm/s. In the deeper atmosphere (${\sim}3000-6000$ hPa), updrafts near the water clouds measure less than 1 cm/s, and decrease with pressure. In our model, this is the large (grid) scale vertical velocity rather than the velocity of individually convecting clouds. These  are diagnosed below using CAPE/CIN.

\begin{figure*}
    \includegraphics[width=\textwidth]{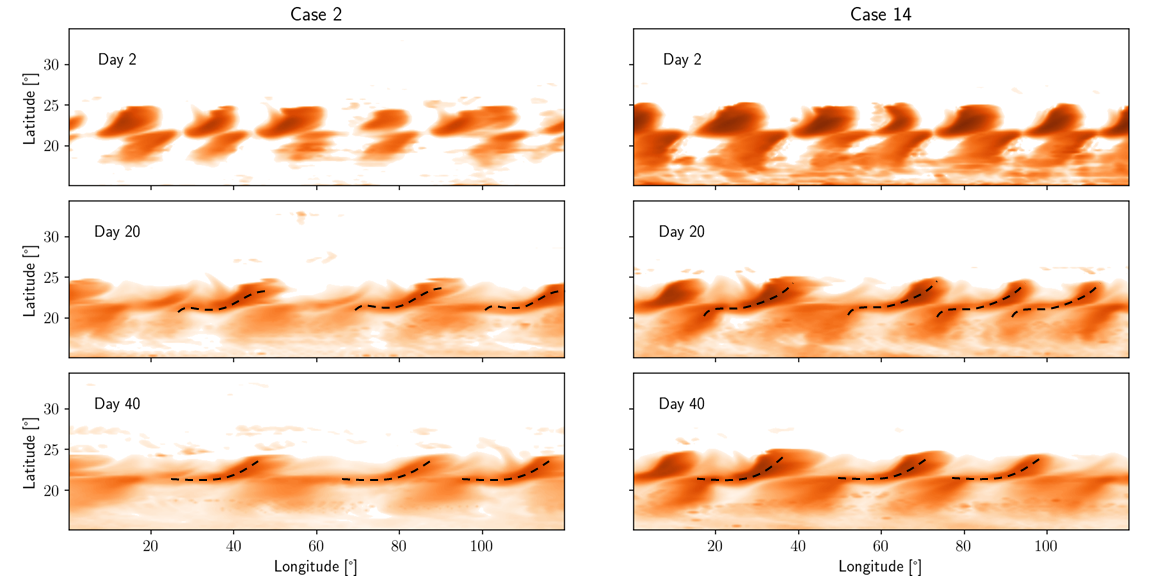}
    \caption{Vertically integrated ammonia solid cloud density for Cases 2 ($2\times$ solar ammonia) and 14 ($4\times$ solar ammonia) at days 2, 20 and 40. The black dashed lines denote the location of the chevron-shaped clouds in our model, which are similar to those observed post plume outbreak on Jupiter. }
    \label{fig:ammonia_evolve}
\end{figure*}

\subsection{Ammonia cloud evolution}
Figure~\ref{fig:ammonia_evolve} shows the evolution of ammonia clouds for Case 2 and Case 14. Ammonia clouds initially form only over the jet, and eventually form a thin layer to the south of the jet. The northern flank is clear for the duration of the simulation. Ammonia clouds generally form above 700 hPa, but do not have a clearly defined base (note the lack of clear bases on Figure~\ref{fig:wave_slice}, as compared to the water clouds), but are rather transient and dynamically coupled to the flow of the wave. In this way, they are akin to dense cirrus-like clouds on Earth, which is consistent with the analysis of \citet{Carlson1988}. Within the jet, ammonia clouds form periodic chevron-shaped features. 

\begin{figure}
    \centering
    \includegraphics[width=\columnwidth]{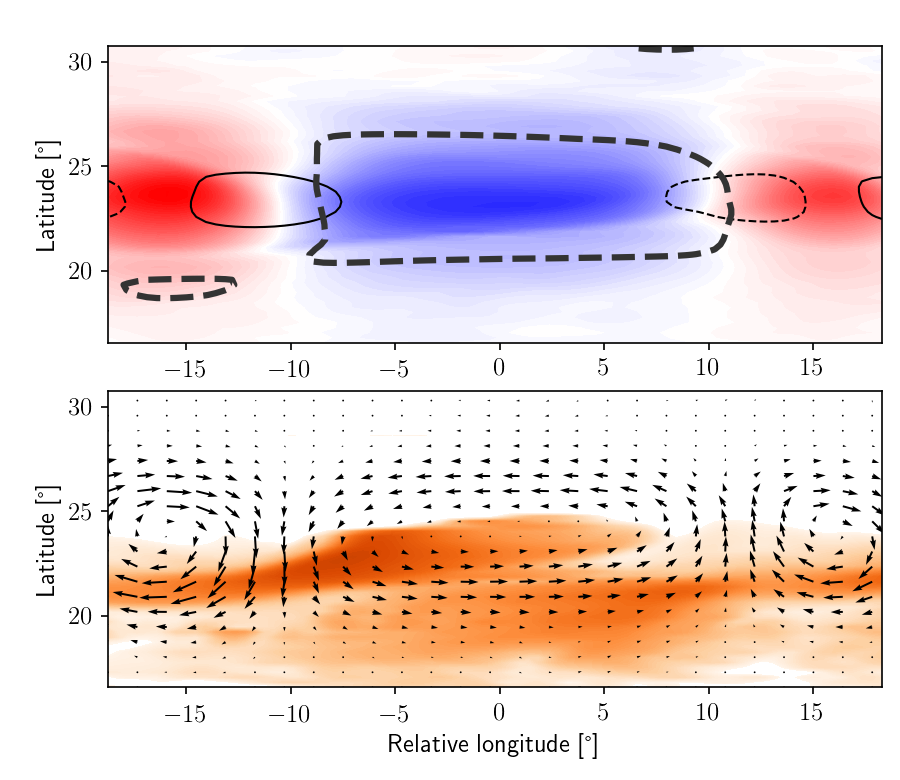}
    \caption{Top: Thermal anomaly at 400 hPa is shown in color (red being warmer/blue being cooler) compared to the zonal mean. The cyclonic wave packet is shown with the thick dashed line, while the thin black lines show vertical velocity (solid for upwelling and dashed for downwelling). Bottom: Integrated ammonia ice cloud density with eddy circulation at 400 hPa. The clouds form within the cooler atmosphere within the cyclonic wave, but the chevron shape is formed by the anticyclonic circulation advecting the cloud to form a west-facing tail.}
    \label{fig:ammonia_wave}
\end{figure}

The ammonia cloud layer is the most directly affected due to the close vertical proximity to the wave packets. Figure~\ref{fig:ammonia_wave} shows the temperature difference below the cyclonic packet and the corresponding integrated cloud density and eddy velocity. 
The cloud formation is triggered by the combination of cooling as a result of the cyclonic anomaly and upwelling of ammonia vapor from depth. The condensed ammonia is pulled into the chevron shape before being pushed to the deeper atmosphere by the downdrafts where it evaporates. These chevron shapes are traced by the dashed line in Figure~\ref{fig:ammonia_evolve}.

Ammonia clouds are maintained by this circulation, and most of the clouds which form within the anticyclonic packets evaporate quickly in the subsaturated atmosphere. As a result, the ammonia clouds are very uniform throughout the simulations, and do not deviate from their chevron shape in the vicinity of the wave. In this regard, the ammonia clouds act as a tracer for the PV packets, since the shape of the cloud is strongly tied to the location of either the cyclonic or anticyclonic packet.



\begin{figure*}
    \includegraphics[width=\textwidth]{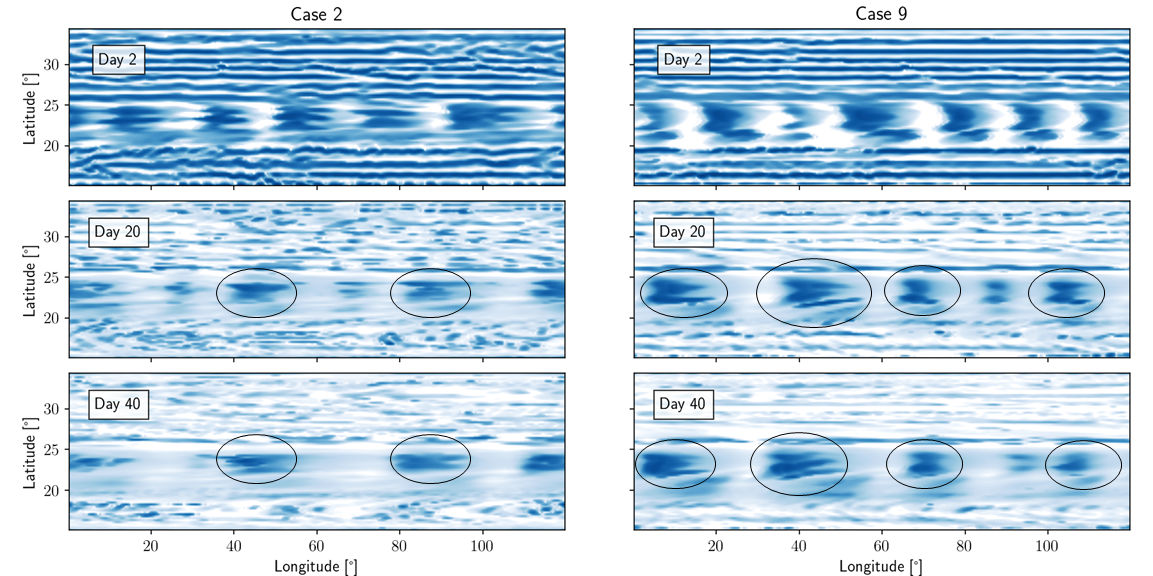}
    \caption{Vertically integrated water solid cloud density for Cases 2 ($1\times$ solar water) and 9 ($2\times$ solar water) at days 2, 20 and 40. The black circles show the location of the thick water clouds that form directly below the cyclonic wavepacket, under the chevron-shaped ammonia clouds. The distinc chevron shape is lacking here. }
    \label{fig:water_evolve}
\end{figure*}
\subsection{Water cloud evolution}
Figure~\ref{fig:water_evolve} shows the evolution of water clouds in our model for Cases 2 and 9, which form below the ammonia layer shown in Figure~\ref{fig:ammonia_evolve} -- here we do not plot the ammonia layer, so as to see the deeper water layer. Water clouds are initially global and thick, with some zonal banding. The base of water clouds are between 4-5 bars, depending on the initial deep abundance value, with the 0.5$\times$ solar case forming higher up. Over time, precipitation and falling cloud particles deplete the clouds in regions where vertical updrafts do not replenish the water vapour. 

\begin{figure}
    \centering
    \includegraphics[width=\columnwidth]{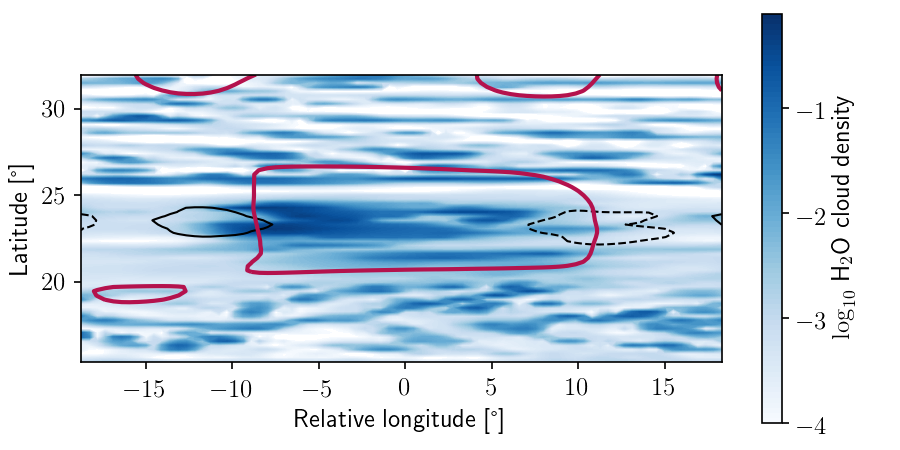}
    \caption{Vertically integrated water solid cloud density with vertical velocity at 4000 hPa in black countours (upwelling denoted by solid) and positive PV anomaly shown in red. The chevron structure of the ammonia clouds is distinctly lacking here. }
    \label{fig:water_wave}
\end{figure}

Figure~\ref{fig:water_wave} shows a closeup of the water clouds below cyclonic wave packet with the vertical velocity at the 4 bar level. Here, unlike the ammonia layer, water vapor is homogeneously mixed both to the north and south of the jet, and eddy velocities are much smaller. As a result, the distinct chevron shapes in the ammonia layer are lacking here, but the presence of the upper level wave can be clearly seen, as thick water clouds form directly below the wave, as a result of the dynamic orography of the wave packets. The clouds form immediately to the east of the updrafts and disappear with the downdrafts.

Flakes of water clouds (`scooters') break off from the thick region below the cyclonic anomaly and are advected by the wind. Due to the amount of cloud material and the region being generally saturated, the broken clouds persist beyond the wave packets until they precipitate or evaporate. This makes the water cloud layer less uniform and more chaotic than the ammonia layer. 

\subsection{Terminal velocity and dissipation of cloud particles}
From our microphysics scheme, we can diagnose the particle sizes and terminal velocity, in a similar fashion to \citet{Hadland2020}, who did the same for methane clouds on Neptune. In the case of ice clouds, we assume that each grid cell has an average cloud particle size. Therefore, we determine this by first calculating the average particle number density to calculate the mass of each particle, and then calculate the diameter from the particle mass, as given in \citet{Palotai2008}. Consequently, we can determine the terminal velocity of the cloud particles in the grid cell using the average particle size. 
 To precipitate as snow, the ice clouds must reach a critical diameter, at which point the terminal velocity for snow is calculated assuming hexagonal plate-like particles. On Jupiter, these snow particles do not generally melt to form rain due to the low temperatures. Instead, snow particles sublimate when they reach the subsaturated atmosphere, returning the mass back to vapor form in the deeper atmosphere.

Below the cyclonic packet, ammonia cloud densities are on the order of $10^{-6}$ kg/m$^3$. The thickest clouds contain particles with a maximum diameter of ${\sim}300-400\mu$m. These largest particles fall with a terminal velocity of between $40-50$ cm/s in the thin upper atmosphere. However, the average terminal velocity of clouds in the jet is an order of magnitude lower at $5-6$ cm/s. In the ammonia cloud layer, terminal velocity is slower than the updraft velocity at the west end of the cyclonic wavepacket, and therefore, new particles are lofted up easily and maintained by the updrafts within the cyclonic packet. 

Ammonia snow only forms in the first few hours, after which there is insufficient ammonia cloud mass to reach the critical diameter for snow to form, as stated above. Therefore, the ammonia clouds instead dissipate as a result of the temperature changes and advection of dry air by the downdrafts. That is to say, ammonia clouds are extremely transient in our simulations, and form and dissipate over short timescales, since these clouds are unable to accrete enough material to form precipitation. Such clouds have been seen on Jupiter, appearing and disappearing over short timescale \citep[e.g.,][]{Choi2013}.

Water clouds have a mean density of $10^{-4}$ kg/m$^3$, approximately two orders of magnitude greater than ammonia clouds. The particles below the center of the wave packet sizes grow beyond the critical diameter to precipitate and convert to snow, and the average diameter of the cloud particles is ${\sim}10\mu$m. The largest cloud particles are near the center of wave and fall at ${\sim}56$ m/s, while those near the edge of the wave (${\sim}\pm5\degree$ longitude in Figure~\ref{fig:water_wave}) are smaller and fall at around $4-6$ cm/s. Regardless, the timescale to fall one scale height, even for the largest cloud particles, is more than $13$ hours. Therefore, the clouds do not precipitate fully in the time that they are advected from one end of the wave packet to the other. In the other regions however, the precipitation dissipates the cloud quickly over a few days. 

Water snow forms throughout the simulation within the water cloud layer with a maximum density of about $10^{-4}$ kg/m$^3$. Any snow that falls below the cloud base ends up in air that is too warm and dry for snow to remain, resulting in quick evaporation \citep{Palotai2008}. The bulk terminal velocity of snow within a grid cell is calculated in the model, and this has a maximum of $1-2$ m/s. On average, snow within the jet falls at about $3-5$ cm/s. Since this is faster than the updraft velocity at the 4 bar level, snow is an effective mechanism to dissipate cloud particles. As stated above, these updrafts correspond to large scale vertical motion in our model. Convective updrafts from individual clouds will loft these snow particles to the upper atmosphere, but explicit treatment of this mechanism will be part of a future study. Similarly to the ammonia clouds, this dissipation does not occur within a single wave packet. Consequently, some water clouds that form within the wave are sometimes advected far beyond the wave packet before dissipating, making water clouds much more long-lived features compared to the ammonia clouds.



\begin{figure*}
    \centering
    \includegraphics[width=\textwidth]{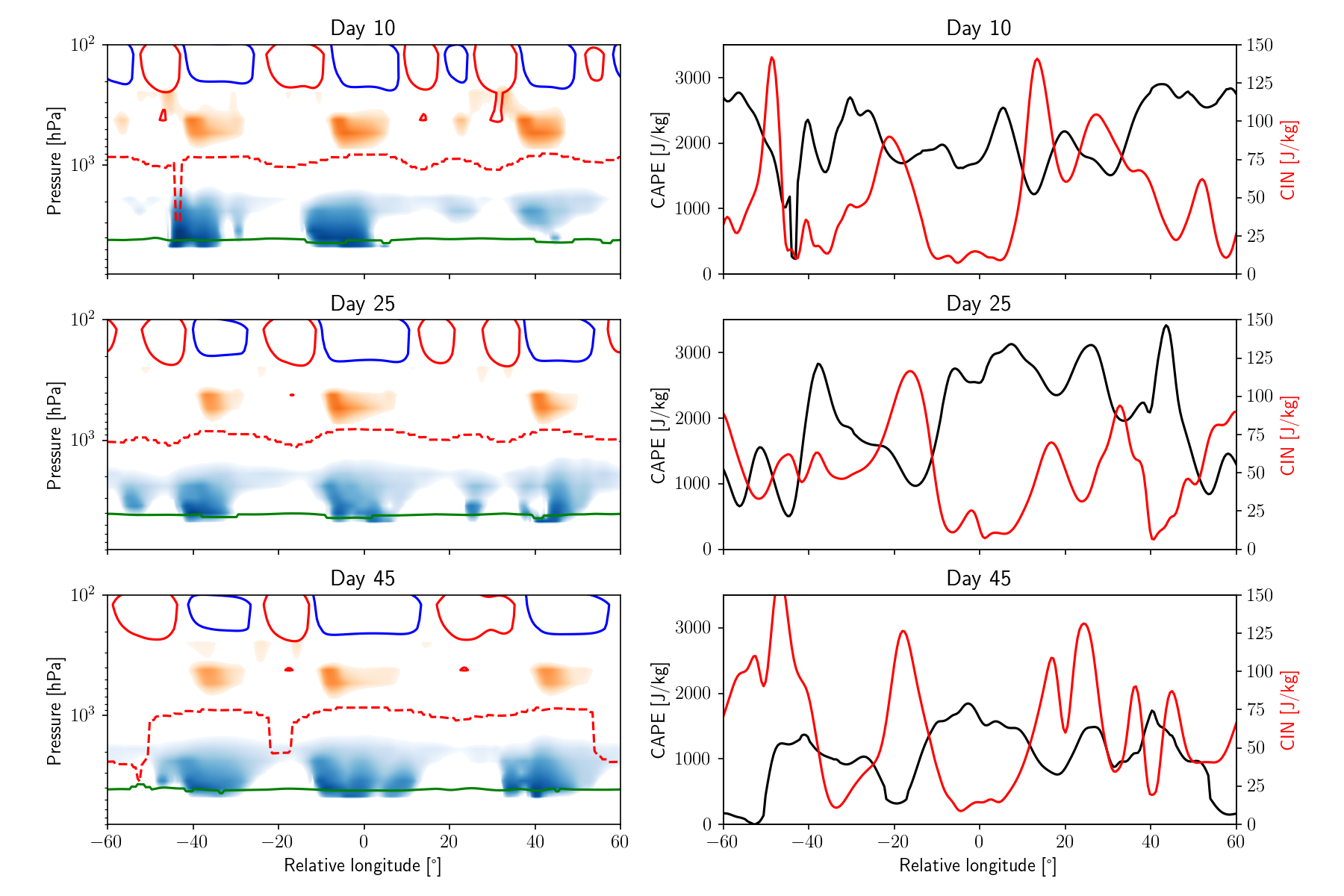}
    \caption{Meridional slice at $23.7\degree$ latitude for Case 2. For each day, the left panel shows ammonia clouds in orange and water clouds in blue. The upper level eddy PV is shown in solid blue for cyclonic anomaly and solid red for anticyclonic. The dashed red is the equilibrium level (EL) and green is the level of free convection (LFC). The right panels show the corresponding CAPE (black) and CIN (red). 
    The longitude axis in each frame is measured relative to the center of the same cyclonic wavepacket. In essence, each panel is moving in the frame of the cyclonic wavepacket.
    The atmosphere below the cyclonic packet shows an increase in CAPE and a decrease in CIN, and vice-versa for the anticylonic packet. The ELs are also characteristically higher below the cyclonic packet, showing that these regions support the trigger of deep convection. }
    \label{fig:CAPE_case2}
\end{figure*}

\subsection{CAPE and diagnosis for convection}
Figure~\ref{fig:CAPE_case2} shows the evolution of CAPE determined from an initial parcel location of 5 bars for Case 2 in frame travelling with the wave. Here, both water and ammonia are enriched below the cyclonic wave packet, as is the CAPE. Below the anticyclonic packet, CAPE is diminished and CIN is increased. This is due to the height of convective region (between the LFC and EL -- green and dashed red line, respectively, in Figure~\ref{fig:CAPE_case2}) is also increased below the cyclonic packet and reduced below the anticyclonic.

On Earth, the analysis of CAPE has three major starting location, the surface (SBCAPE), which uses the values at the surface, mean layer (MLCAPE), which takes an average of the region between the surface and cloud base, and most unstable (MUCAPE), which provides the CAPE of the most unstable parcel in the sounding. These different values provide an idea of the range of convective potential in that sounding, with MLCAPE generally being the smallest, to MUCAPE being the largest.  However, in our model, given the very coarse vertical resolution, it is difficult to explicitly differentiate between these three different ideologies, as successive vertical layers will have significantly different volatile abundances.  To that end, starting a layer too deep results in a large region of negative buoyancy, while starting too high up results in very little gain in convective potential.
We find that $4$ bars is too high to be used as a starting pressure, as it is usually within the clouds, and $6$ bars is too deep and results in either the parcel convecting too strongly, or not at all. 
Therefore, for all further analysis, we start the parcel at $5$ bars when determining the CAPE/CIN and related quantities.


CAPE is primarily affected by the water mixing ratio at the parcel's initial location and vertical sounding of the atmosphere. An increase in CAPE is generally a result of an increase in deep water content and also a cooler upper atmosphere, and vice versa. CIN is similarly affected: a warm layer below the LFC will result in higher CIN while a cool layer will decrease it. 

\begin{figure}
    \centering
    \includegraphics[width=\columnwidth]{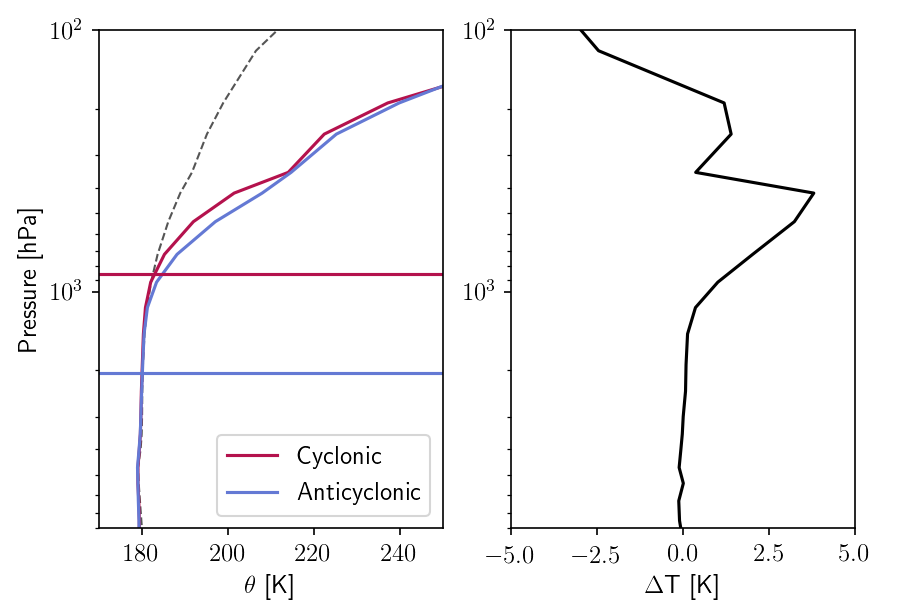}
    \caption{Left: Vertical potential temperature profile of the atmosphere at day 45 below the cyclonic in red and below the anticyclonic in blue. These points correspond to $0\degree$ and $-20\degree$ longitude, respectively, on the third panel of Figure~\ref{fig:CAPE_case2}. The horizontal lines show the location of the equilibrium level for a parcel starting at 5 bar. The potential temperature profile of this parcel is shown in the thin gray dashed line. Right: Temperature difference between the anticylonic and the cyclonic showing the increase in upper level temperatures below the anticyclonic packet. }
    \label{fig:CAPE_sounding}
\end{figure}

Figure~\ref{fig:CAPE_sounding} shows the potential temperature structure below the cyclonic and anticylonic packets for day 45 of Case 2. In our model, the lower troposphere is naturally convective because of the adiabatic extension to the temperature-pressure profile below 6 bars, resulting in nearly constant potential temperature profile near the water cloud layer. As a result, there is a base value of CAPE throughout our model, even in regions that are dry and warm. 

In the upper troposphere, the baroclinic wave affects the upper level temperatures, leading to a hotter atmosphere below the anticylonic packet and cooler one below the cyclonic. Therefore, below the cyclonic packet, an upwelling parcel will retain its buoyancy due to being warmer than the cold atmosphere, resulting in the taller convective region. The opposite occurs below the anticylonic packet, resulting in a shallower convection and more convective inhibition.

For the deep convective water plumes to occur, we require a sounding with high CAPE and low CIN. The parcel must then have enough vertical velocity to counter the CIN and reach the LFC. Therefore, in this region, convective plumes will occur where CIN is low and there is sufficient positive vertical velocity. In our model, this occurs to the west of the cyclonic packets. 

The value of CAPE in the atmosphere is generally below ${\sim}~3000$ J/kg, while the value of CIN is below ${\sim}~100$ J/kg. Since this is the energy per unit mass that parcel gains from latent heat release (or requires to convect in the case of CIN), we can determine the maximum vertical velocity of a convecting parcel by equating its vertical kinetic energy to the CAPE/CIN, as,

\begin{equation}
    \dfrac{1}{2}w_{\rm max}^2 {\sim} CAPE \Rightarrow w_{\rm max} {\sim} \sqrt{2\times CAPE},
\end{equation}
and similarly for CIN. Therefore, to break the CIN of ${\sim} 100$ J/kg, the parcel will need a vertical velocity of about 14 m/s below the anticylonic packets. Below the cyclonic packets, CIN is negligible, and thus, the upwelling will reach a peak velocity of about $70$ m/s with a CAPE of ${\sim} 3000$ J/kg.  Mean updraft velocities are generally half as strong as the peak value, and thus, for a CAPE value of 3000 J/kg, the mean velocity is about 30 m/s.  
We are currently adding a sub-grid scale moist convective parameterization to the EPIC model, with which we will be able to resolve these updraft velocities.


\subsection{Volatile abundance}

\begin{figure}
    \centering
    \includegraphics[width=\columnwidth]{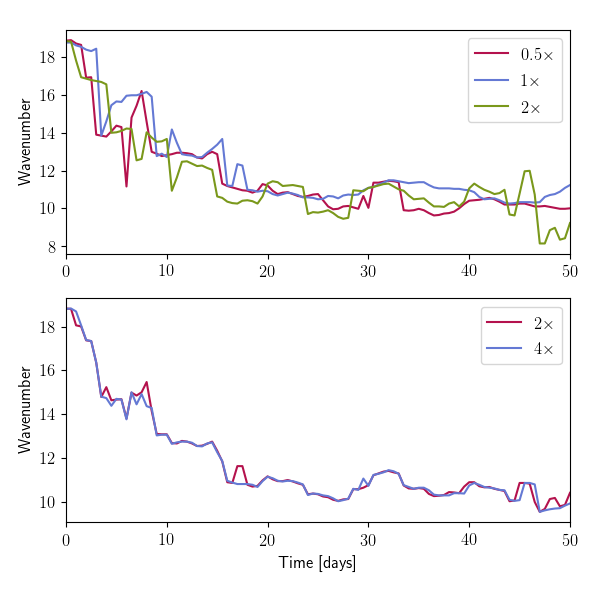}
    \caption{Evolution of wavenumber for different water abundances (top) and ammonia abundance (bottom) as a function of model time. These values are determined by averaging over the different simulations with different perturbation profiles for the same deep abundance. There is no significant difference in the dynamics of the waves due to variation in volatile abundance.}
    \label{fig:wave_water}
\end{figure}

Figure~\ref{fig:wave_water} shows the evolution of the wavenumber for different water and ammonia abundances. The amount of condensibles does not seem to have an effect on the dynamics of the wave, likely due to the latter being in a pressure level where both species are negligible. 

Convection is primarily driven by the wave but convective ability is strongly dependent on the available water. As such, it serves as a valuable tracer to probe the deep water abundance. In our model, we tested three different water abundances: $0.5\times$, $1\times$ and $2\times$ the nominal solar [O/H] ratio. 

Figure~\ref{fig:ammonia_evolve} shows the distinction between a $2\times$ and $4\times$ solar deep abundance on the resulting clouds. In both cases, the north of the jet (NTrB) is void of ammonia clouds, but the clouds in the $4\times$ solar case are thicker and extend further south. 

Figure~\ref{fig:water_evolve} shows the difference between $1\times$ solar and $2\times$ solar water on the evolution of the water clouds in the region. The higher water abundance leads to denser clouds in general, as expected, and these persist throughout the duration of the simulation. The higher abundance also results in more clouds forming to the north and south of the jet, within the two neighbouring belts. However, morphologically speaking, there is no effect of varying the water abundances on the shape and structure of the cloud; the cyclonic packet guides the locations for cloud formation and the anticyclonic packet results in cloud clearing. 

\begin{figure}
    \centering
    \includegraphics[width=\columnwidth]{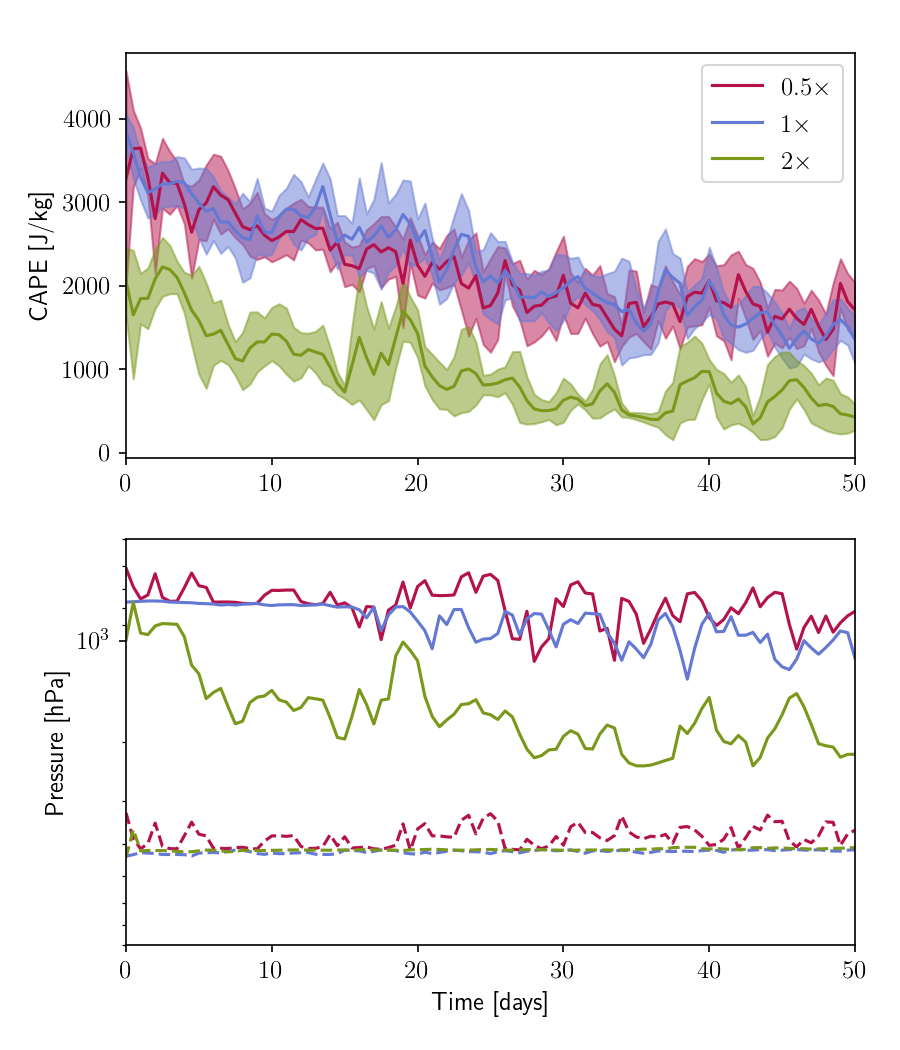}
    \caption{Top: Evolution of CAPE for different water abundances. All parcels here start at 5 bars. Bottom: LFC (dashed) and EL (solid) as a function of time in the model for different water abundances.}
    \label{fig:CAPE_abundance}
\end{figure}

Figure~\ref{fig:CAPE_abundance} shows the time evolution of CAPE for different abundances averaged over all perturbations and ammonia abundances, with the shaded region showing the $1\sigma$ confidence in the average. The bottom panel shows the average LFC and EL for a parcel starting at 5 bars. The $0.5\times$ and $1\times$ solar abundances are nearly equal but the $2\times$ solar case has a significantly lower CAPE value. In our spinup cases, we noticed that the $2\times$ solar cases generally have a lower temperature at the water cloud base compared to other two, while the $0.5\times$ solar is slightly higher. Therefore, the added weight of the water, combined with changes in $c_p$ result in a difference in the dynamical steady-state of the atmosphere in our model. Accordingly, the parcel in a $2\times$ solar atmosphere loses buoyancy quickly and thus has a smaller value of CAPE even though it has the same LFC as in the $1\times$ solar cases. As such, we find that increasing the water abundance does not proportionately increase the convective potential; rather, the opposite effect is seen here. In this case, the major contribution is actually from the thermal structure of the atmosphere, as the differences in CAPE are a result of the cooling/heating in the deeper atmosphere.

\section{Discussion}
\label{sec:discussion}

In this work, we have detailed our simulations of Jupiter's 24$\degree$ N jet with the EPIC model to study the effect of atmospheric dynamics on cloud features, and vice versa. We investigated the effects of varying the deep abundance of water and ammonia and perturbed the atmosphere with random vertices. We found that a  baroclinic instability forms above $200$ hPa, which results in a potential vorticity wave. The motion of the wave results in up- and downwelling which strongly influences the formation of both water and ammonia clouds. Initially the wavenumber is ${\sim}16-18$ but drops over the duration of the simulation to ${\sim}8-11$. 

A similar phenomenon was observed near the Equatorial Belt (EB) by \citet{Choi2013}, who noticed a wave of hot spots and plumes travelling at ${\sim}100$ m/s. The ammonia clouds formed a similar chevron shape due to the circulation within the wave, and several small `scooter' clouds were advected by the zonal wind faster than the wave, and disappeared into the hot spots and reappeared at the plumes. The waves measured by \citet{Choi2013} were much larger than we observe in our model, and we attribute this to the differences in the region of study, as the jet stream in the EB is much wider than the one at $24\degree$ N. There are many aspects of this region in our model that are seen in observations, but also point to various discrepancies in our assumptions. In the following subsections, we discuss these discrepancies in the context of prior studies of this region.

\subsection{Observed ammonia cloud structure}

\begin{figure}
    \centering
    \includegraphics[width=\columnwidth]{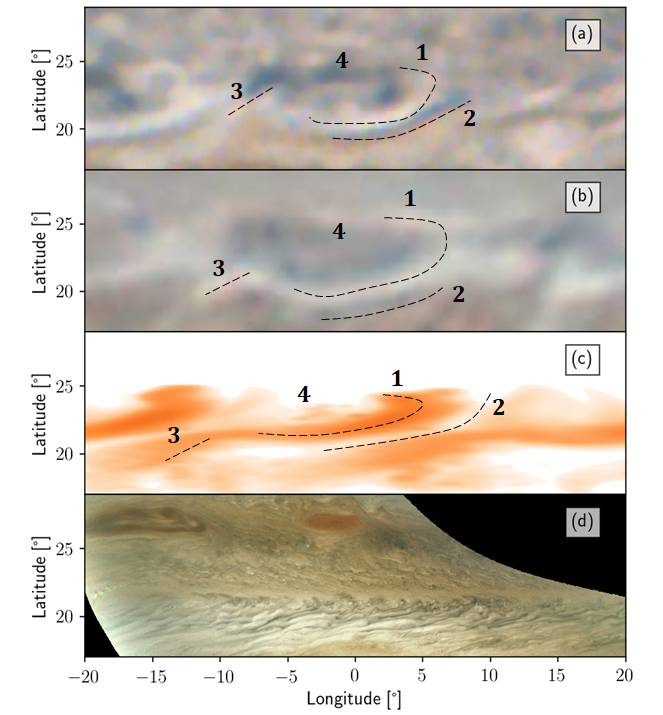}
    \caption{Images showing the modeled region, all set to the same physical scale. (a) Observation of the chevron clouds in the wake of the 2016 outbreak (Adapted from \citet{SanchezLavega2017}; Credit: JL Dauvergne/Fran\c{c}ois Colas/S2P/IMCCE/OMP). (b) Similar clouds seen after the 2020 outbreak (Credit: Clyde Foster/ASSA/PVOL). (c) Model output from Case 5 showing the integrated ammonia cloud density and similar chevron shape at day 7.5. (d) {\it JunoCam} image from Perijove 22 showing the quiescent state with waves of much smaller wavelength. In (a), (b) and (c), 1 represents the bright ammonia chevron feature, 2 represents the adjacent cloud clearing, 3 represents the tail, and the 4 is the central cloud clearing within the anticyclonic wave packet. All four of these features are matched in our model (c). In the {\it JunoCam} image (d), these large scale structures are not seen, but small scale waves are ubiquitous. }
    \label{fig:obs_comp_model}
\end{figure}

The wave in our model has an associated thermal anomaly, with the air being ${\sim}1.5-2$ K cooler below the cyclonic anomaly and warmer by the same amount below the anticyclonic packet. This in turn affects the ammonia vapor humidity, and thus the dry, warm anticyclonic packet is cloud free, while the cool, moist cyclonic packet contains thick ammonia clouds. Upwelling occurs to the west of the cyclonic packet, and thus brings in more vapor from below and the eddy circulation results in chevron shaped waves. These chevron features are similar to those seen by \citet{SanchezLavega2017} in the wake of the water plume outbreak in the jet in 2016. They were unable to reproduce these features in potential vorticity maps when they modeled the region with a dry atmosphere in EPIC, but concluded that the clouds are the result of anticylonic circulation within the jet. Our simulations are able to reproduce the size ($10\degree {\sim} 11,000$ km) and location of these features (Figure~\ref{fig:obs_comp_model}), but hint at a more complicated circulation pattern: the clouds are a result of upwelling between the cyclonic and anticyclonic packets (Figure~\ref{fig:ammonia_wave}). The ``head", or the eastern end of the chevron, forms within the cool cyclonic packet, while the ``tail", or the westward end, is a result of anticyclonic circulation to the west of the updraft. Both ends disappear when they meet the warm core of the anticyclonic packet, where the downdrafts result in evaporation of the cloud particles. 

In our simulations, the wave forms as a natural consequence of perturbing this region, and persists regardless of physical parameters (vapor abundance and perturbation profile). The shape, structure and speed of the wave are independent of the nature of the random vortex initialization and volatile abundance, and thus, we interpret this wave to form naturally in the jovian atmosphere when perturbed, e.g., through plumes in nature, which are emulated by vortices in our model.
Over time, turbulent mixing removes this feature and results in a uniform zonal structure \citep{SanchezLavega2017}. Indeed, observations of Jupiter during the quiescent periods between outbreaks do not reveal these large (${\sim}10\degree$ diameter) waves. Instead, the jet is riddled with similar features but on a significantly smaller scale (e.g., as in Figure~\ref{fig:obs_comp_model}d), which require very fine resolution to resolve. Such resolution has been possible with images from {\it Cassini}, {\it New Horizons}, {\it Hubble} and more recently {\it Juno}. The modeled ammonia clouds do not resolve this fine structure, as the model resolution is too coarse,  but do show the wave dissipating over time and the chevron clouds stretching out zonally. It is possible that our model is either too low resolution, or too dissipative to retain the evolution of these waves on the smaller scale. Furthermore, the transition between the NTB and NTZ has several persistent large vortices, which influences these waves. We currently do not add any such vortices and thus, are neglecting their effect on the clouds. Lastly, since the wave forms a scale height above the ammonia cloud base, it is possible that the wave continues to exist in the upper atmosphere, but cannot be traced as the ammonia cloud precipitates downwards after a plume outbreak. The wave likely reappears following the next upwelling when the volatiles are pushed up to the pressure level where the wave forms.

\subsection{Drift rate and vertical wind shear}

Since these chevrons travel with the wave, the apparent velocity of these clouds is the same as that of the wave (${\sim}75$ m/s), even though the zonal wind velocity at the cloud level is much higher (${\sim}120$-$140$ m/s in our model). Indeed, \citet{SanchezLavega2017} found that the chevron clouds travel much slower than the jet velocity (at ${\sim}100$ m/s) even within the peak of the jet. Therefore, using cloud tracking to determine these the cloud top velocities produces a much lower velocity than present in the atmosphere at that pressure level. 
Similarly to \citet{Choi2013}, we caution that in regions where thermal anomalies dictate cloud formation, and individual clouds (like the `scooters') cannot be fully resolved, cloud tracking would likely result in an inaccurate measurement of the wind speeds in the region. Upper tropospheric wind velocities are difficult to quantify, however, as there are no visible targets to track. One possible avenue is to use Doppler shifts from spectral lines, similar to \citet{Goncalves2019}, who use a spectrometer to obtain Doppler wind speeds measured in the visible spectrum. Their method would be able to sense actual particle velocities, but they note that the limited spectral and spatial resolution degrades the accuracy of the data near regions where the dynamics deviate significantly from the zonal mean, e.g. near vortices. 

\begin{figure}
    \centering
    \includegraphics[width=\columnwidth]{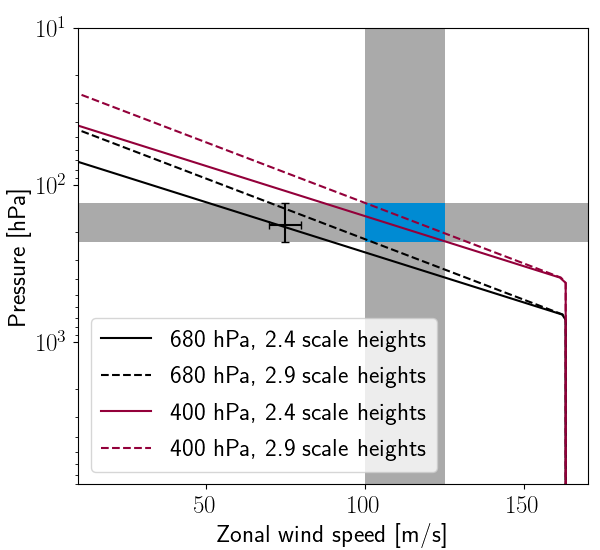}
    \caption{Comparison of wind decay rate \citep[as determined by ][]{Gierasch1986} and observed cloud top pressure on the wave speed. The vertical shaded region represents the measured speed of the wave by \citet{SanchezLavega2017}, while the horizontal shaded region is the typical center pressure of the wave in our model. The black point shows the location and speed of the center of the wave in our model, with corresponding errors. The blue shaded region combines the constraints of both the observed velocity of the wave determined by \citet{SanchezLavega2017} and vertical location of the wave, and this is best matched by moving the pressure of observed wind speeds to $400$ hPa (red lines), from $680$ hPa level (black lines). }
    \label{fig:wind_shear}
\end{figure}

On the other hand, the fact that these clouds can be used as a tracer for a region of the atmosphere where clouds do not form presents an opportunity to constrain the wind shear profile above the cloud tops. 
Two possible scenarios exist -- if we track the cloud tops from images, the wind velocity determined could correspond to a lower pressure than the assumed value of $680$ hPa as the cloud tops can exists as high as $400$ hPa, as shown in our simulation. Therefore, the zonal wind decay is accordingly shifted to higher altitudes, and the mean zonal wind velocity at ${\sim}150$ hPa is likely closer to $100$ m/s (Figure~\ref{fig:wind_shear}). Alternatively, we could be seeing a slower wind decay above the clouds, implying the wind extends further into the stratosphere. Indeed, \citet{Gierasch1986} found that the decay rate near the $24\degree$ N jet was about $2.9$ scale heights, which indicates a slower decay compared to the global average of $2.4$ scale heights that we use in our model. However, the slower decay rate is not enough to explain the much higher wave speeds measured by \citet{SanchezLavega2017} (Figure~\ref{fig:wind_shear}), given the vertical location of the wave in our model. Moving the start of the decay up to $400$ hPa provides a much better match to observed wave speed and modeled wave location. However, even so, we are unable to differentiate between the wind decay rate, and thus the parameters that affect the formation and dynamics of this wave will need to be investigated thoroughly to resolve this discrepancy.  


\subsection{Water clouds and abundance}

Shifting the focus to the deeper, water layer, the thermal anomaly at this pressure level in our model is negligible, but updrafts are still present at the west edge of the cyclonic packet. These updrafts are on the order of $1$ cm/s and result in water clouds forming directly below the cyclonic wave packet. These water clouds have a less uniform shape compared to the ammonia and also form thick `scooter' clouds, which persist well beyond the cyclonic packet when advected by the zonal flow. 

We vary the abundance of water in the deep atmosphere between $0.5\times$, $1\times$ and $2\times$ the nominal solar [O/H] ratio and $2\times$ and $4\times$ the nominal [N/H] ratio to investigate the effect on the dynamics of the wave and resulting cloud formation. 
The addition of volatiles affects the molar mass, which affects the dynamics of large scale features, \citep[e.g.,][]{Hadland2020}. However, in this case, the wave forms above the cloud, in a region where the added vapor mass is insignificant, the dynamics of the wave is unaffected by the amount of added volatiles.

We have so far ignored the contributions from the NH$_4$SH cloud deck. While this species may be responsible for observed features in the visible cloud deck, it is unlikely to result in large scale changes in the dynamics of this region, as the cloud density for NH$_4$SH is comparable to ammonia clouds \citep{Carlson1988}. With a $3\times$ the solar [S/H], the density of NH$_4$SH clouds is an order of magnitude smaller than water \citep{AtreyaWong1999}, with a lower latent heat release, and therefore is likely not a significant contributor to the convective potential of the atmosphere.

As such, the wave affects the clouds, but not the other way around. Outside the wave, the effect of abundance is more apparent: The higher abundances, for both species, generally result in thicker clouds and more persistent cloud formation in the belt to the south of the jet. These are expected since we assume a uniform, well mixed region below the cloud base. Consequently, the cool belt to the south is more saturated with higher abundances for both species. 

\subsection{Convection and CAPE}

For the different cases, we are primarily interested in the convective potential and inhibition. Due to the hydrostatic nature of EPIC, we simply diagnose the onset of convection and its strength rather than physically modeling convective events. We find that the thermal anomaly as a result of the wave packets is a strong driver for convection as the cyclonic packets experience an increase in CAPE and reduction in CIN, while the anticylonic packets demonstrate the opposite. This relation holds for all parameters tested in our model. Indeed, the region to the east of the cyclonic packets act similar to hot spots and form a dry, convectively stable layer through the mechanism suggested by \citet{ShowmanDowling2000}, where the relative motion of the warm anticylonic packet compresses the column of air downward and leads to a subsaturated atmosphere and cloud evaporation. Conversely, below the cyclonic packet, we see the opposite -- the cold air advection results in expansion of the deeper atmosphere and an increase in saturation and CAPE \citep{DowlingGierasch1989}. CAPE in the model peaks at ${\sim}3000$ J/kg, which corresponds to a peak convective updraft velocity of ${\sim}70$ m/s, which is similar to the model results of \citet{Stoker1986} and \citet{Hueso2001}. We do not see the high updraft velocities of $w>100$ m/s due to lower initial relative humidity in our model. 

The effects of increasing water mass on convection have been analyzed before \citep{Guillot1995,Li2015,Leconte2017}, and these studies have placed a critical limit of $10\times$ solar water, beyond which the weight of water limits the convective potential and results in convective storms never being triggered. On the other hand, \citet{Li2015} specify that a mixing ratio between $2-5\times$ solar is not significant enough to cause a direct mass loading effect, which is larger than the range of values we test here. Accordingly, we do not see a direct effect of water mass loading on CAPE/CIN. Instead, CAPE for different water abundances varied in our model as a result of the change in the dynamical steady state that formed as a consequence of adding water. Specifically, the $0.5\times$ solar was warmer and the $2\times$ solar case was cooler at the water cloud base, compared to the $1\times$ solar at the end of the 2D adjustment phase. We interpret this to be due to the non-linearity of thermal (e.g., changes in heat capacity) and mechanical (mass-loading) effects of adding/removing water, coupled with the dynamics. These facets of water convection are difficult to relate with theory, and thus, require the use of 3-dimensional numerical models. 

Ultimately, analysis of CAPE in this study is simply a measure of whether convection can occur, and estimating the relative strength of convective storms. We will do a follow-up study of this phenomenon using a moist convective module that will effectively resolve cloud formation in a convective setting. It is clear, however, that the addition of water does not lead to a simple relation for diagnosing convection, and so our future analyses will span the range of water abundances to better understand this link. 

\section{Conclusions}
We have simulated the region between $15-33.5\degree$ N planetographic latitude using our EPIC model, mainly focusing on ammonia and water cloud formation within the $24\degree$ N jet. We perturb the region using vortices and study the resulting cloud formation in the context of convective ability. We determine the CAPE and CIN due to water condensation, starting with a reference parcel at 4 bars. 

We find that a persistent upper level (${\sim}180$ hPa) vorticity wave has a profound effect on the cloud structure. In our model, the wave has a wavenumber between ${\sim}8-11$. The cyclonic packets are cooler and are punctuated by up- and downwelling on the west and east ends respectively. The upwelling enriches volatiles from the deeper layers and results in thick water and ammonia cloud formation below the cyclonic packet. The enrichment, combined with the cooler upper levels result in a relative increase in CAPE in this region. Conversely, below the anticyclonic packet, downwelling and heating bring in dry air and evaporates both cloud decks, and also leads to a reduction in CAPE and increase in CIN. 

\section*{Data Availability}
The data underlying this article will be shared on request to the corresponding author.

\acknowledgements
We acknowledge support by the NASA's Early Career Fellowship (Grant No. 80NSSC18K0183), NASA's Solar System Workings (Grants No. NNX16AQ0), NASA's Cassini Data Analysis (Grant No. 80NSSC19K0198) and Future Investigators in NASA Earth and Space Science and Technology (Grant No. 80NSSC19K1541) programs. We thank Dr. Steven Lazarus for the insightful discussions on PV waves, and Dr. Timothy Dowling for his help with the bookkeeping in the EPIC model.

\appendix

\section{CAPE scheme}
\label{appendix:CAPE}
When a moist parcel is forced up, its temperature is initially defined by a dry adiabatic expansion until the parcel becomes saturated at a level defined as the lifting condensation level (LCL). Following this, further ascent causes the parcel to condense, releasing latent heat. This diabatic heat release causes the parcel to warm up and it will now follow a moist adiabatic ascent where the temperature of parcel decreases slower than when it was dry. When the parcel is warmer than the surrounding atmosphere, it can now freely convect due to being buoyant, defined as the level of free convection (LFC). Eventually the parcel becomes cooler than the atmosphere (e.g. near the tropopause), and thus becomes negatively buoyant, at the equilibrium level (EL). 

Thus, to study moist-convection, the above process needs to be quantified. Earth-based meteorology tackles this issue by determining the convective available potential energy (CAPE) in a region. CAPE is the energy per unit mass a parcel of air gains from moist-convective (i.e. latent-heat release driven) ascent. It is obtained by integrating the total positive buoyant force experienced by the parcel. The parcel is buoyant between the LFC and EL, and thus the CAPE is given by,

\begin{equation}
    \label{eq:CAPE}
    {\rm CAPE} = \int_{\rm LFC}^{\rm EL} \dfrac{F}{m} dz = - R_d \int_{\rm LFC}^{\rm EL} (T_{\rm v} - T'_{\rm v}) d \ln P,
\end{equation}
where the dry specific gas constant is $R_d$. $T_{\rm v}$ is the virtual temperature of the atmosphere and $T'_{\rm v}$ is the virtual temperature of the parcel, which is defined as,

\begin{equation}
    T_{\rm v} = T \dfrac{q + \epsilon}{\epsilon(1 + q)},
\end{equation}

\noindent where $q$ is the mass mixing ratio of the condensing species and $\epsilon$ is the ratio of molar masses of the moist to dry vapor. On Jupiter, for water condensation, $\epsilon=18/2.2\approx8.2$.

Conversely, it is useful to define the barrier that a moist parcel has to cross to be able to convect freely. This is the energy that a parcel must have to reach the LFC, and is defined as the total negative buoyant force experienced by the parcel between its initial level and the LFC. This is the convective inhibition (CIN), given by,

\begin{equation}
    \label{eq:CIN}
    \text{CIN} = -R_d \int_{p_0}^\text{LFC} (T_\text{v} - T'_\text{v}) d \ln P .
\end{equation}
The negative sign is added to make the value positive for convenience.

The definition of CAPE involves integrating a parcel of air from a chosen pressure level up through the atmosphere. Since model data are coarse, the thermodynamic variables are interpolated onto a finer grid, which allows for sub-gridcell determination of the LFC and EL. The buoyancy (integrand in the CAPE) is calculated on the fine grid, and interpolation is done numerically between the two points. In summary, the calculation of CAPE is done in the following manner:
\begin{enumerate}
    \item Determine the LCL by raising the parcel on a dry adiabatic until saturated. 
    \item Construct the moist parcel ascent from the LCL up through the atmosphere on the coarse grid
    \item Interpolate the sounding onto the fine grid and calculate the buoyancy (integrand in Eqns. \ref{eq:CAPE} and \ref{eq:CIN})
    \item Determine the LFC and EL on the fine grid, as the first zero points on the buoyancy. 
    \item Integrate the buoyancy to find CAPE and CIN
\end{enumerate}



\bibliographystyle{aasjournal}
\bibliography{ref}{}

\begin{thebibliography}{}
\expandafter\ifx\csname natexlab\endcsname\relax\def\natexlab#1{#1}\fi
\providecommand{\url}[1]{\href{#1}{#1}}
\providecommand{\dodoi}[1]{doi:~\href{http://doi.org/#1}{\nolinkurl{#1}}}
\providecommand{\doeprint}[1]{\href{http://ascl.net/#1}{\nolinkurl{http://ascl.net/#1}}}
\providecommand{\doarXiv}[1]{\href{https://arxiv.org/abs/#1}{\nolinkurl{https://arxiv.org/abs/#1}}}

\bibitem[{{Atreya} {et~al.}(1999){Atreya}, {Wong}, {Owen}, {Mahaffy},
  {Niemann}, {de Pater}, {Drossart}, \& {Encrenaz}}]{AtreyaWong1999}
{Atreya}, S.~K., {Wong}, M.~H., {Owen}, T.~C., {et~al.} 1999, \planss, 47,
  1243, \dodoi{10.1016/S0032-0633(99)00047-1}

\bibitem[{{Bjoraker}(2020)}]{Bjoraker2020}
{Bjoraker}, G.~L. 2020, Nature Astronomy, 4, 558,
  \dodoi{10.1038/s41550-020-1086-3}

\bibitem[{{Bjoraker} {et~al.}(2015){Bjoraker}, {Wong}, {de Pater}, \&
  {{\'A}d{\'a}mkovics}}]{Bjoraker2015}
{Bjoraker}, G.~L., {Wong}, M.~H., {de Pater}, I., \& {{\'A}d{\'a}mkovics}, M.
  2015, \apj, 810, 122, \dodoi{10.1088/0004-637X/810/2/122}

\bibitem[{{Bjoraker} {et~al.}(2018){Bjoraker}, {Wong}, {de Pater}, {Hewagama},
  {{\'A}d{\'a}mkovics}, \& {Orton}}]{Bjoraker2018}
{Bjoraker}, G.~L., {Wong}, M.~H., {de Pater}, I., {et~al.} 2018, \aj, 156, 101,
  \dodoi{10.3847/1538-3881/aad186}

\bibitem[{{Carlson} {et~al.}(1988){Carlson}, {Rossow}, \&
  {Orton}}]{Carlson1988}
{Carlson}, B.~E., {Rossow}, W.~B., \& {Orton}, G.~S. 1988, Journal of
  Atmospheric Sciences, 45, 2066,
  \dodoi{10.1175/1520-0469(1988)045<2066:CMOTGP>2.0.CO;2}

\bibitem[{{Charney} \& {Stern}(1962)}]{CharneyStern}
{Charney}, J.~G., \& {Stern}, M.~E. 1962, Journal of Atmospheric Sciences, 19,
  159, \dodoi{10.1175/1520-0469(1962)0190159:OTSOIB2.0.CO;2}

\bibitem[{{Choi} {et~al.}(2013){Choi}, {Showman}, {Vasavada}, \&
  {Simon-Miller}}]{Choi2013}
{Choi}, D.~S., {Showman}, A.~P., {Vasavada}, A.~R., \& {Simon-Miller}, A.~A.
  2013, \icarus, 223, 832, \dodoi{10.1016/j.icarus.2013.02.001}

\bibitem[{{Cunningham}(1910)}]{Cunningham1910}
{Cunningham}, E. 1910, Proceedings of the Royal Society of London Series A, 83,
  357, \dodoi{10.1098/rspa.1910.0024}

\bibitem[{{Dowling} \& {Gierasch}(1989)}]{DowlingGierasch1989}
{Dowling}, T.~E., \& {Gierasch}, P.~J. 1989, in Bulletin of the American
  Astronomical Society, Vol.~21, 946

\bibitem[{{Dowling} {et~al.}(2006){Dowling}, {Bradley}, {Col{\'o}n}, {Kramer},
  {LeBeau}, {Lee}, {Mattox}, {Morales-Juberias}, {Palotai}, {Parimi}, \&
  {Showman}}]{Dowling2006}
{Dowling}, T.~E., {Bradley}, M.~E., {Col{\'o}n}, E., {et~al.} 2006, \icarus,
  182, 259, \dodoi{10.1016/j.icarus.2006.01.003}

\bibitem[{{Fletcher}(2017)}]{Fletcher2017b}
{Fletcher}, L.~N. 2017, \grl, 44, 4725, \dodoi{10.1002/2017GL073806}

\bibitem[{{Fletcher} {et~al.}(2017){Fletcher}, {Orton}, {Rogers}, {Giles},
  {Payne}, {Irwin}, \& {Vedovato}}]{Fletcher2017}
{Fletcher}, L.~N., {Orton}, G.~S., {Rogers}, J.~H., {et~al.} 2017, \icarus,
  286, 94, \dodoi{10.1016/j.icarus.2017.01.001}

\bibitem[{{Fletcher} {et~al.}(2020){Fletcher}, {Orton}, {Greathouse}, {Rogers},
  {Zhang}, {Oyafuso}, {Eichst{\"a}dt}, {Melin}, {Li}, {Levin}, {Bolton},
  {Janssen}, {Mettig}, {Grassi}, {Mura}, \& {Adriani}}]{Fletcher2020}
{Fletcher}, L.~N., {Orton}, G.~S., {Greathouse}, T.~K., {et~al.} 2020, Journal
  of Geophysical Research (Planets), 125, e06399, \dodoi{10.1029/2020JE006399}

\bibitem[{{Garc{\'\i}a-Melendo} {et~al.}(2005){Garc{\'\i}a-Melendo},
  {S{\'a}nchez-Lavega}, \& {Dowling}}]{GarciaMelendo2005}
{Garc{\'\i}a-Melendo}, E., {S{\'a}nchez-Lavega}, A., \& {Dowling}, T.~E. 2005,
  \icarus, 176, 272, \dodoi{10.1016/j.icarus.2005.02.012}

\bibitem[{{Gierasch} {et~al.}(1986){Gierasch}, {Conrath}, \&
  {Magalh{\~a}es}}]{Gierasch1986}
{Gierasch}, P.~J., {Conrath}, B.~J., \& {Magalh{\~a}es}, J.~A. 1986, \icarus,
  67, 456, \dodoi{10.1016/0019-1035(86)90125-9}

\bibitem[{{Gierasch} {et~al.}(2000){Gierasch}, {Ingersoll}, {Banfield},
  {Ewald}, {Helfenstein}, {Simon-Miller}, {Vasavada}, {Breneman}, {Senske}, \&
  {Galileo Imaging Team}}]{Gierasch2000}
{Gierasch}, P.~J., {Ingersoll}, A.~P., {Banfield}, D., {et~al.} 2000, \nat,
  403, 628, \dodoi{10.1038/35001017}

\bibitem[{{Gon{\c{c}}alves} {et~al.}(2019){Gon{\c{c}}alves}, {Schmider},
  {Gaulme}, {Morales-Juber{\'\i}as}, {Guillot}, {Rivet}, {Appourchaux},
  {Boumier}, {Jackiewicz}, {Sato}, {Ida}, {Ikoma}, {M{\'e}karnia}, {Underwood},
  \& {Voelz}}]{Goncalves2019}
{Gon{\c{c}}alves}, I., {Schmider}, F.~X., {Gaulme}, P., {et~al.} 2019, \icarus,
  319, 795, \dodoi{10.1016/j.icarus.2018.10.019}

\bibitem[{{Guillot}(1995)}]{Guillot1995}
{Guillot}, T. 1995, Science, 269, 1697, \dodoi{10.1126/science.7569896}

\bibitem[{{Hadland} {et~al.}(2020){Hadland}, {Sankar}, {LeBeau}, \&
  {Palotai}}]{Hadland2020}
{Hadland}, N., {Sankar}, R., {LeBeau}, Raymond~Paul, J., \& {Palotai}, C. 2020,
  \mnras, 496, 4760, \dodoi{10.1093/mnras/staa1799}

\bibitem[{Holton(2004)}]{Holtonbook}
Holton, J.~R. 2004, {An introduction to dynamic meteorology}, 4th edn., ed.
  R.~Dmowska \& J.~R. Holton, International Geophysics Series (Burlington, MA:
  Elsevier Academic Press,), 535.
\newblock \url{http://books.google.com/books?id=fhW5oDv3EPsC}

\bibitem[{{Hong} {et~al.}(2004){Hong}, {Dudhia}, \& {Chen}}]{HDC2004}
{Hong}, S.-Y., {Dudhia}, J., \& {Chen}, S.-H. 2004, Monthly Weather Review,
  132, 103, \dodoi{10.1175/1520-0493(2004)132<0103:ARATIM>2.0.CO;2}

\bibitem[{{Hueso} \& {S{\'a}nchez-Lavega}(2001)}]{Hueso2001}
{Hueso}, R., \& {S{\'a}nchez-Lavega}, A. 2001, \icarus, 151, 257,
  \dodoi{10.1006/icar.2000.6606}

\bibitem[{{Hueso} {et~al.}(2002){Hueso}, {S{\'a}nchez-Lavega}, \&
  {Guillot}}]{Hueso2002}
{Hueso}, R., {S{\'a}nchez-Lavega}, A., \& {Guillot}, T. 2002, Journal of
  Geophysical Research (Planets), 107, 5075, \dodoi{10.1029/2001JE001839}

\bibitem[{{Ingersoll} {et~al.}(2000){Ingersoll}, {Gierasch}, {Banfield},
  {Vasavada}, \& {Galileo Imaging Team}}]{Ingersoll2000}
{Ingersoll}, A.~P., {Gierasch}, P.~J., {Banfield}, D., {Vasavada}, A.~R., \&
  {Galileo Imaging Team}. 2000, \nat, 403, 630, \dodoi{10.1038/35001021}

\bibitem[{{Leconte} {et~al.}(2017){Leconte}, {Selsis}, {Hersant}, \&
  {Guillot}}]{Leconte2017}
{Leconte}, J., {Selsis}, F., {Hersant}, F., \& {Guillot}, T. 2017, \aap, 598,
  A98, \dodoi{10.1051/0004-6361/201629140}

\bibitem[{{Li} \& {Ingersoll}(2015)}]{Li2015}
{Li}, C., \& {Ingersoll}, A.~P. 2015, Nature Geoscience, 8, 398,
  \dodoi{10.1038/ngeo2405}

\bibitem[{{Li} {et~al.}(2017){Li}, {Ingersoll}, {Janssen}, {Levin}, {Bolton},
  {Adumitroaie}, {Allison}, {Arballo}, {Bellotti}, {Brown}, {Ewald}, {Jewell},
  {Misra}, {Orton}, {Oyafuso}, {Steffes}, \& {Williamson}}]{Li2017}
{Li}, C., {Ingersoll}, A., {Janssen}, M., {et~al.} 2017, \grl, 44, 5317,
  \dodoi{10.1002/2017GL073159}

\bibitem[{{Li} {et~al.}(2020){Li}, {Ingersoll}, {Bolton}, {Levin}, {Janssen},
  {Atreya}, {Lunine}, {Steffes}, {Brown}, {Guillot}, {Allison}, {Arballo},
  {Bellotti}, {Adumitroaie}, {Gulkis}, {Hodges}, {Li}, {Misra}, {Orton},
  {Oyafuso}, {Santos-Costa}, {Waite}, \& {Zhang}}]{Li2020}
{Li}, C., {Ingersoll}, A., {Bolton}, S., {et~al.} 2020, Nature Astronomy, 4,
  609, \dodoi{10.1038/s41550-020-1009-3}

\bibitem[{{Limaye}(1986)}]{Limaye1986}
{Limaye}, S.~S. 1986, \icarus, 65, 335, \dodoi{10.1016/0019-1035(86)90142-9}

\bibitem[{{Morales-Juber{\'\i}as} {et~al.}(2011){Morales-Juber{\'\i}as},
  {Sayanagi}, {Dowling}, \& {Ingersoll}}]{MoralesJuberias2011}
{Morales-Juber{\'\i}as}, R., {Sayanagi}, K.~M., {Dowling}, T.~E., \&
  {Ingersoll}, A.~P. 2011, \icarus, 211, 1284,
  \dodoi{10.1016/j.icarus.2010.11.006}

\bibitem[{{Moses} {et~al.}(2005){Moses}, {Fouchet}, {Bezard}, {Gladstone},
  {Lellouch}, \& {Feuchtgruber}}]{Moses2005}
{Moses}, J.~I., {Fouchet}, T., {Bezard}, B., {et~al.} 2005, Journal of
  Geophysical Research (Planets), 110, E08001, \dodoi{10.1029/2005JE002411}

\bibitem[{{Niemann} {et~al.}(1998){Niemann}, {Atreya}, {Carignan}, {Donahue},
  {Haberman}, {Harpold}, {Hartle}, {Hunten}, {Kasprzak}, {Mahaffy}, {Owen}, \&
  {Way}}]{GalileoProbepaper}
{Niemann}, H.~B., {Atreya}, S.~K., {Carignan}, G.~R., {et~al.} 1998, \jgr, 103,
  22831, \dodoi{10.1029/98JE01050}

\bibitem[{{Palotai} \& {Dowling}(2008)}]{Palotai2008}
{Palotai}, C., \& {Dowling}, T.~E. 2008, \icarus, 194, 303,
  \dodoi{10.1016/j.icarus.2007.10.025}

\bibitem[{{Palotai} {et~al.}(2014){Palotai}, {Dowling}, \&
  {Fletcher}}]{Palotai2014}
{Palotai}, C., {Dowling}, T.~E., \& {Fletcher}, L.~N. 2014, \icarus, 232, 141,
  \dodoi{10.1016/j.icarus.2014.01.005}

\bibitem[{{P{\'e}rez-Hoyos} {et~al.}(2020){P{\'e}rez-Hoyos},
  {S{\'a}nchez-Lavega}, {Sanz-Requena}, {Barrado-Izagirre},
  {Carri{\'o}n-Gonz{\'a}lez}, {Anguiano-Arteaga}, {Irwin}, \&
  {Braude}}]{PerezHoyos2020}
{P{\'e}rez-Hoyos}, S., {S{\'a}nchez-Lavega}, A., {Sanz-Requena}, J.~F.,
  {et~al.} 2020, \icarus, 352, 114031, \dodoi{10.1016/j.icarus.2020.114031}

\bibitem[{Pruppacher \& Klett(2010)}]{PruppacherKlettbookCh10}
Pruppacher, H., \& Klett, J. 2010, Hydrodynamics of Single Cloud and
  Precipitation Particles (Dordrecht: Springer Netherlands), 361--446,
  \dodoi{10.1007/978-0-306-48100-0_10}

\bibitem[{{Rogers}(2019)}]{Rogers2019}
{Rogers}, J.~H. 2019, Journal of the British Astronomical Association, 129, 13

\bibitem[{{S{\'a}nchez-Lavega} {et~al.}(2008){S{\'a}nchez-Lavega}, {Orton},
  {Hueso}, {Garcia-Melendo}, {Perez-Hoyos}, {Simon-Miller}, {Rojas},
  {G{\'o}mez}, {Yanamandra-Fisher}, {Fletcher}, {Joels}, {Kemerer}, {Hora},
  {Karkoschka}, {de Pater}, {Wong}, {Marcus}, {Pinilla-Alonso}, {Carvalho},
  {Go}, {Parker}, {Salway}, {Valimberti}, {Wesley}, \&
  {Pujic}}]{SanchezLavega2008}
{S{\'a}nchez-Lavega}, A., {Orton}, G.~S., {Hueso}, R., {et~al.} 2008, \nat,
  451, 437, \dodoi{10.1038/nature06533}

\bibitem[{{S{\'a}nchez-Lavega} {et~al.}(2017){S{\'a}nchez-Lavega}, {Rogers},
  {Orton}, {Garcia-Melendo}, {Legarreta}, {Colas}, {Dauvergne}, {Hueso},
  {Rojas}, {Perez-Hoyos}, {Mendikoa}, {I{\~n}urrigarro}, {Gomez-Forrellad},
  {Momary}, {Hansen}, {Eichstaedt}, {Miles}, \& {Wesley}}]{SanchezLavega2017}
{S{\'a}nchez-Lavega}, A., {Rogers}, J.~H., {Orton}, G.~S., {et~al.} 2017, \grl,
  44, 4679, \dodoi{10.1002/2017GL073421}

\bibitem[{{Sayanagi} {et~al.}(2010){Sayanagi}, {Morales-Juber{\'\i}as}, \&
  {Ingersoll}}]{Sayanagi2010}
{Sayanagi}, K.~M., {Morales-Juber{\'\i}as}, R., \& {Ingersoll}, A.~P. 2010,
  Journal of Atmospheric Sciences, 67, 2658, \dodoi{10.1175/2010JAS3315.1}

\bibitem[{{Showman}(2007)}]{Showman2007}
{Showman}, A.~P. 2007, Journal of Atmospheric Sciences, 64, 3132,
  \dodoi{10.1175/JAS4007.1}

\bibitem[{{Showman} \& {Dowling}(2000)}]{ShowmanDowling2000}
{Showman}, A.~P., \& {Dowling}, T.~E. 2000, Science, 289, 1737,
  \dodoi{10.1126/science.289.5485.1737}

\bibitem[{{Stoker}(1986)}]{Stoker1986}
{Stoker}, C.~R. 1986, \icarus, 67, 106, \dodoi{10.1016/0019-1035(86)90179-X}

\bibitem[{{Stratman} {et~al.}(2001){Stratman}, {Showman}, {Dowling}, \&
  {Sromovsky}}]{Stratman2001}
{Stratman}, P.~W., {Showman}, A.~P., {Dowling}, T.~E., \& {Sromovsky}, L.~A.
  2001, \icarus, 151, 275, \dodoi{10.1006/icar.2001.6603}

\bibitem[{{Taylor} {et~al.}(2004){Taylor}, {Atreya}, {Encrenaz}, {Hunten},
  {Irwin}, \& {Owen}}]{Taylor2004}
{Taylor}, F.~W., {Atreya}, S.~K., {Encrenaz}, T., {et~al.} 2004, {The
  composition of the atmosphere of Jupiter}, ed. F.~{Bagenal}, T.~E. {Dowling},
  \& W.~B. {McKinnon}, Vol.~1 (Cambridge University Press), 59--78

\bibitem[{{Thompson} {et~al.}(2007){Thompson}, {Mead}, \&
  {Edwards}}]{Thompson2007}
{Thompson}, R.~L., {Mead}, C.~M., \& {Edwards}, R. 2007, Weather and
  Forecasting, 22, 102, \dodoi{10.1175/WAF969.1}

\bibitem[{{Wong} {et~al.}(2004){Wong}, {Bjoraker}, {Smith}, {Flasar}, \&
  {Nixon}}]{Wong2004PSS}
{Wong}, M.~H., {Bjoraker}, G.~L., {Smith}, M.~D., {Flasar}, F.~M., \& {Nixon},
  C.~A. 2004, \planss, 52, 385, \dodoi{10.1016/j.pss.2003.06.005}

\bibitem[{Wong {et~al.}(2004)Wong, Mahaffy, Atreya, Niemann, \&
  Owen}]{Wong2004}
Wong, M.~H., Mahaffy, P.~R., Atreya, S.~K., Niemann, H.~B., \& Owen, T.~C.
  2004, Icarus, 171, 153

\end{thebibliography}


\end{document}